\documentclass[fleqn,usenatbib,useAMS]{mnras}
\usepackage{newtxtext,newtxmath}
\usepackage[T1]{fontenc}
\usepackage{ae,aecompl}
\usepackage{longtable}
\usepackage{graphicx}	% Including figure files
\usepackage{amsmath}	% Advanced maths commands
\usepackage{float}
\usepackage{multirow, array}
\usepackage{ulem}
\usepackage{hyperref}
\usepackage[]{xcolor}

\newcommand{\orcid}[1]{\href{https://orcid.org/#1}{\includegraphics[width=10pt]{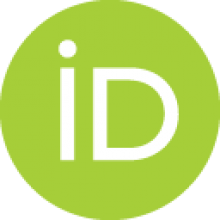}}}

\title[Spectroscopic age indicators in giants]{On the validity of the spectroscopic age indicators [Y/Mg], [Y/Al], [Y/Si], [Y/Ca], and [Y/Ti] for giant stars}

\author[O. J. Katime Santrich et al.]{Orlando J. Katime Santrich$^{1, 2}$\thanks{E-mail: ojksantrich@uesc.br and ojkatime@gmail.com} \orcid{0000-0003-0439-2331}, 
Leandro Kerber$^{1}$\orcid{0000-0002-7435-8748}, Yuri Abuchaim$^{2}$\orcid{0000-0002-6838-2178},
\newauthor{and Geraldo Gon\c calves$^{2}$\orcid{0000-0003-4999-3691}
}
\\
$^{1}$Universidade Estadual de Santa Cruz, UESC, Rodovia Jorge Amado km\,16, Ilh\'eus, 45662000, Bahia, Brazil.\\
$^{2}$Universidade de S\~ao Paulo, Instituto de Astronomia, Geof\'isica e Ci\^encias Atmosf\'ericas, Departamento de Astronomia, \\ Rua do Mat\~ao\,1226, 05508-090 S\~ao Paulo/SP, Brazil.\\
}

\date{Accepted in MNRAS 2022}

\pubyear{2022}

\begin{document}

\label{firstpage}
\pagerange{\pageref{firstpage}--\pageref{lastpage}}
\maketitle

\begin{abstract}
The abundance ratios [Y$/$Mg], [Y$/$Al], [Y$/$Si], [Y$/$Ca], and [Y$/$Ti] have been suggested as chemical clocks for solar-metallicity dwarf stars in the field as well as for giant stars in open clusters. To verify this last hypothesis, we derive these abundances ratios of 50 giant stars belonging to seven open clusters. To calculate the abundances, we analyze FEROS spectra assuming the LTE-hypothesis. We confirm that [Y$/$Mg], [Y$/$Al], [Y$/$Si], [Y$/$Ca], and [Y$/$Ti] work as chemical clocks for field dwarf stars at the local region (d\,$<$\,1\,kpc) whereas for the field giants the [Y$/$Mg], [Y$/$Al] and [Y$/$Si] also present trends with the ages but high scattering. [Y$/$Ca] and [Y$/$Ti] do not present any correlation with ages in the field giants. In Our open clusters, the behaviour is similar, [Y$/$Mg], [Y$/$Al] and [Y$/$Si] present evident trends, whereas [Y$/$Ca]\,vs.\,Ages is a flat and [Y$/$Ti]\,vs.\,Ages is less steep. We also confirm that the chemical clocks have high scatter at the early ages. In the case of the compiled sample, the chemical clocks are similar to our results but in some situations there are important differences. Several relations between abundance ratios and ages may be obtained when dwarfs and giants are analyzed, confirming the non-universality of the spectroscopic age indicators.
\end{abstract}

\begin{keywords}
stars: abundances - stars: fundamental parameters - stars: late-type - Galaxy: evolution - (Galaxy:) open clusters and associations: general
\end{keywords}

\section{Introduction}

The spectroscopic ratios [Y$/$Mg]\footnote{It was obtained from [Y$/$Fe] and [Mg$/$Fe], in turn derived from abundance definition [A$/$B]\,=\,$\log$(A/B)$_{\star}-\log$(A/B)$_{\odot}$.}, and [Y$/$Al]\footnote{from [Y$/$Fe] and [Al$/$Fe].} have been important in last years owing to possibility to obtain reliable ages for the stars. This fact is striking because stellar ages are hard to derive and have always been a topic of discussion in Astrophysics. These spectroscopic clocks have been studied in field dwarf stars, while in giants the results are limited for some open clusters. 

The ratio [Y$/$Mg] has been the most studied spectroscopic indicator in literature mainly for field dwarf stars, being proposed as a promising age indicators for solar twin stars by \citet{niss15} and \citet{TM16}. Afterwards, \citet{fel17} used a sample of 714 dwarf stars in the solar neighbourhood \citep{bensby14} to conclude that [Y$/$Mg]\,vs\,Age relation is unique to solar analogues. \citet{DM19}, using more than 1000 F\,G\,K field dwarfs, confirmed the negative correlation between [Y$/$Mg] and the ages \citep{niss15}. More recently, \citet{ambre} used 342 turn-off stars in the solar neighborhood and also found the same negative tendency but they did not see dependency on the metallicity. \citet{skula} studied [Y$/$Mg] in the dwarf galaxy Sculptur obtaining the same negative inclination with the ages, although their [Y$/$Mg] ratios have presented lower values relative to the Galactic objects. This result was the first to report such clock in other galaxies.

Despite the validation of the ratio [Y/Mg] as a chemical clock for field dwarf stars, the use of this (or any other) spectroscopic age indicator to study the chemical evolution of our Galaxy is limited by the faintness of such stars. The natural solution to probe the distant regions in the Galactic disk are the giant stars, whose luminosity can reach several magnitudes brighter. Although the correlation between abundance ratios and age for giant stars can be verified using those in the Galactic field, it is evident that open clusters are the ideal laboratories for such studies since the stars have similar initial chemical composition and age. In particular, the ages can be very well determined in open clusters by means of isochrone fitting techniques, reaching uncertainties in log(age) of $\sim\,0.01$ if [Fe/H] are known with an accuracy of $\sim 0.06$\,dex \citep{bossini19}, which, translate in linear age, corresponds to just 2$\%$.

The first work dedicated to study the [Y$/$Mg]\,vs\,Age in giants was \citet{slum17} who analyzed five red clump stars in four open clusters (NGC\,6811, NGC\,6819, M\,67, and NCG\,188) ranging ages from $\sim$ 1 Gyr to 6 Gyr. 
They claimed an evident negative tendency between [Y$/$Mg] and their ages, in excellent agreement with the one presented by \citet{niss16} for solar twins. On the other hand, \citet{ps18} also investigated [Y$/$Mg]\,vs\,Age in giant stars in three open clusters younger than $\sim$ 2 Gyr (NGC\,3680, NGC\,2360, and NCG\,5822), concluding that there is no evidence favouring the negative trend. 

The relation [Y$/$Al]\,vs\,Age was studied in first studied by \citet{niss16} who analyzed 21 solar twin stars, \citet{spin18} increased the sample analyzing 79 solar twins. The two works concluded that the ratio [Y$/$Al] is an excellent candidate to be spectroscopic age indicator because of the low scattering shown. Later, \citet{DM19} also confirmed this ratio as age indicator in field dwarf stars. 

\citet{casali} studied the ratios [Y$/$Mg], [Y$/$Al], [Y$/$Si], [Y$/$Ca], [Y$/$TiI] and [Y$/$TiII] in open cluster giants and solar-like stars. These abundance ratios presented a clear scattering for giant stars with ages\,<\,2.0\,Gyr. The authors concluded that this result was due to the variation of the star formation histories at different Galactocentric distances, and to the role of metallicity in the efficiency of the neutron-capture element production.

More recently, \citet{casami21} studied these abundance ratios and the ages for open clusters in the inner and outer disks. They concluded that the tendencies exist but with differences in the two galactic disks.  
This paper is organized as follow. In Section \ref{data}, we present the data employed in this study. In section\,\ref{spectros} we describe spectroscopic techniques needed to calculate the abundance ratios. In Section\,\ref{53}, we present and discuss the spectroscopic clocks. Finally, Section\,\ref{conclusions} shows our conclusions.

\section{Data}\label{data}

We have applied a detailed spectroscopic analysis for 50 giant stars in the open clusters NGC\,5316, NGC\,6633, IC\,4756, NGC\,5822, NGC\,6940, IC\,4651 and NGC\,2682 (M\,67) (Table\,\ref{table_openclusters}). These clusters have solar-like metallicities and ages ranging from 150 Myr to 3.6 Gyr. 

\begin{figure*}
\includegraphics[scale=0.42]{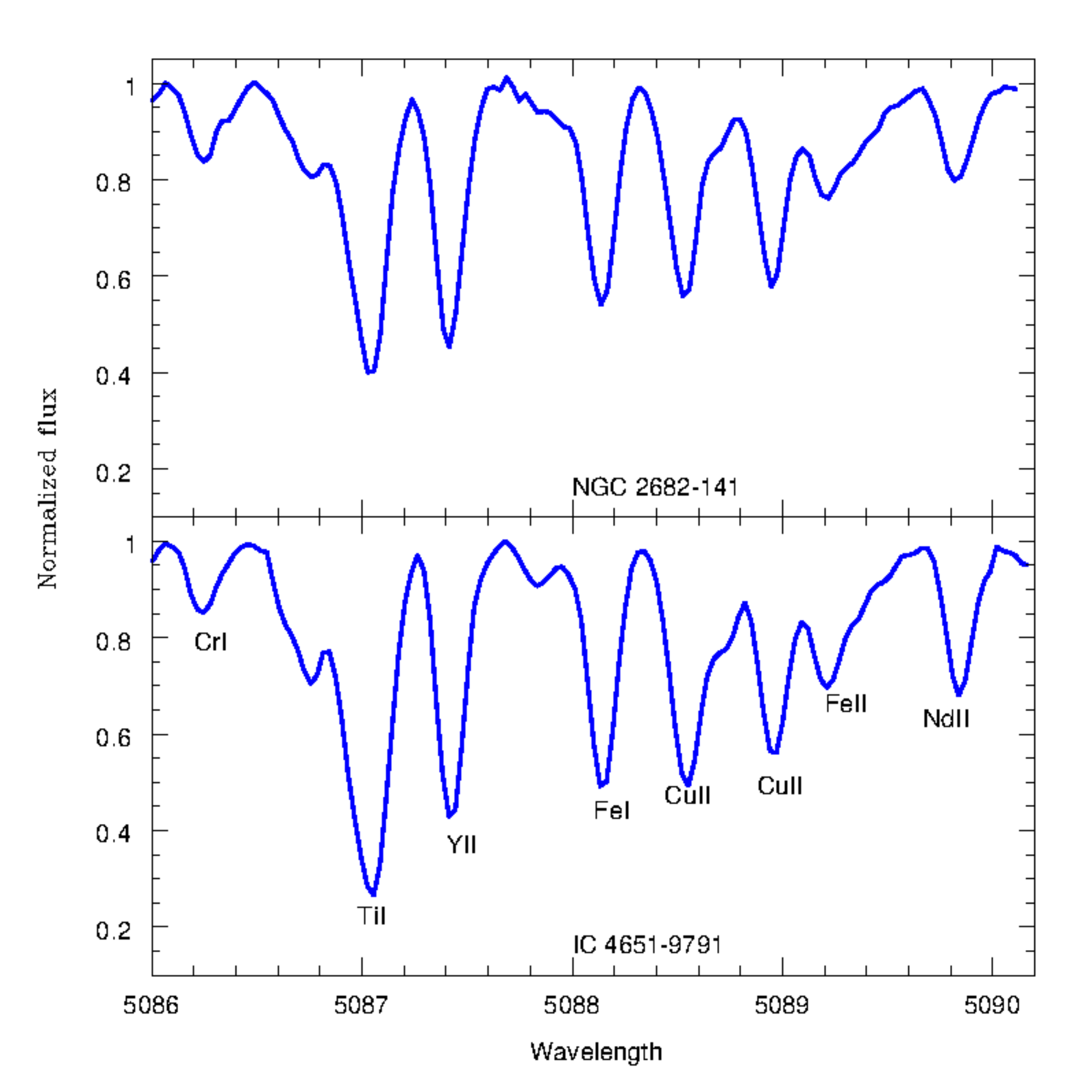}
\includegraphics[scale=0.42]{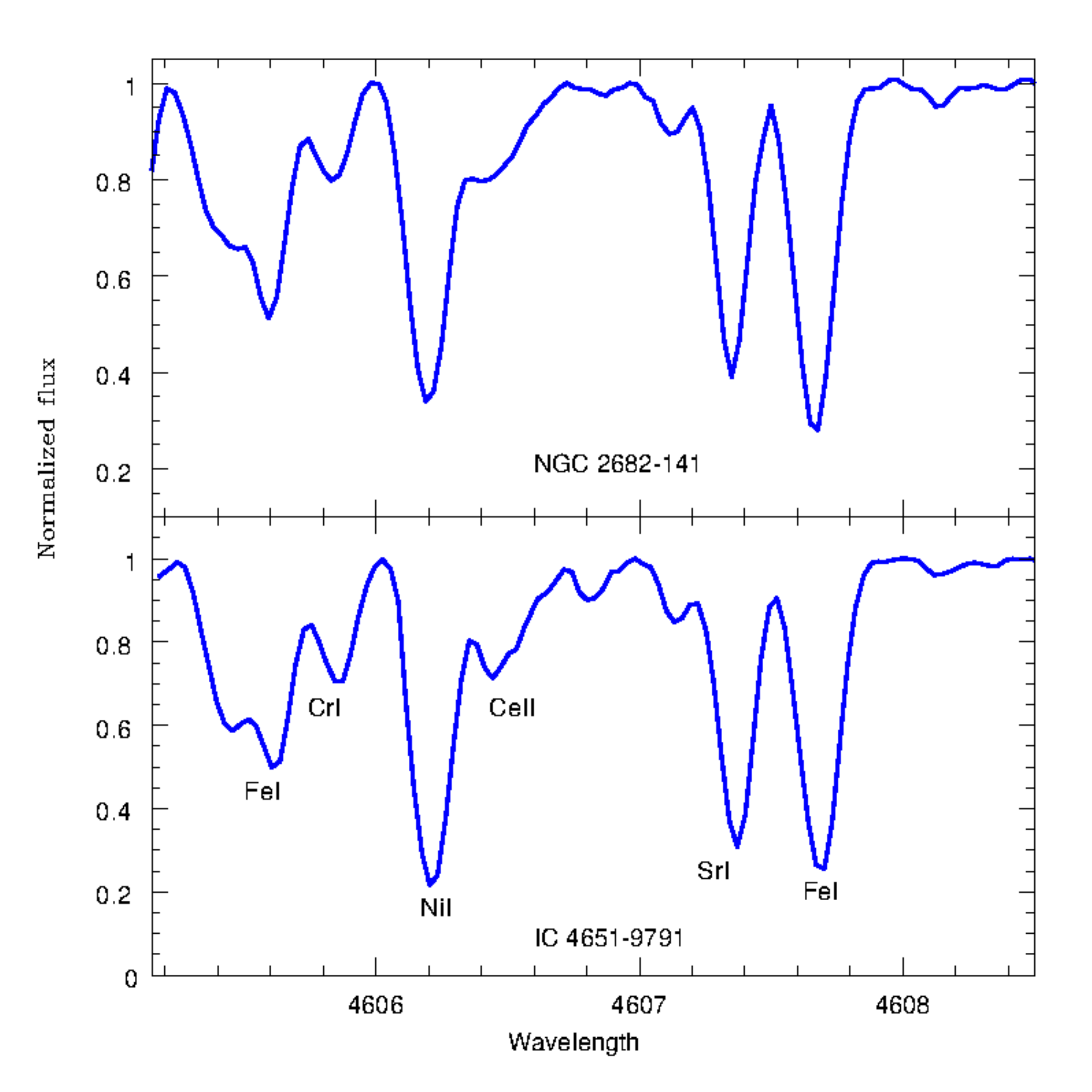}
\caption{Spectral regions around Y\,II and Sr\,I absorption lines in the stars IC\,4651$-$9791 and NGC\,2682$-$141. Some lines of other elements are also identified.}
\label{specreg}
\end{figure*}

In order to calculate the chemical abundances and age indicators, we used high-resolution spectra obtained with the Fiberfed Extended Range Optical Spectrograph 
\citep[FEROS,][]{kauf99}. Its spectral resolving power is R\,=\,48000, corresponding to 2.2 pixels of 15$\mu$, and the wavelength coverage is [3800, 9200]\AA. FEROS is installed in the MPI 2.2\,mts telescope in La Silla$/$Chile. We have taken the reduced spectra from phase\,III form in the ESO$/$archive\footnote{\url{http://archive.eso.org/cms.html}}. The nominal S$/$N ratio was evaluated by measuring the \textit{rms} flux fluctuation in selected continuum windows, and the typical values were S$/$N\,=\,[120, 180]. The Juno solar spectrum used in the analysis has a S$/$N\,=\,800. Therefore, the spectra have good quality to carry out the spectroscopic analysis and to obtain reliable results. This fact is shown in Figure\,\ref{specreg} where Y\,II and Sr\,I absorption lines are identified in two stars of the clusters IC\,4651 and NGC\,2682.

\begin{figure}
\includegraphics[scale=0.48]{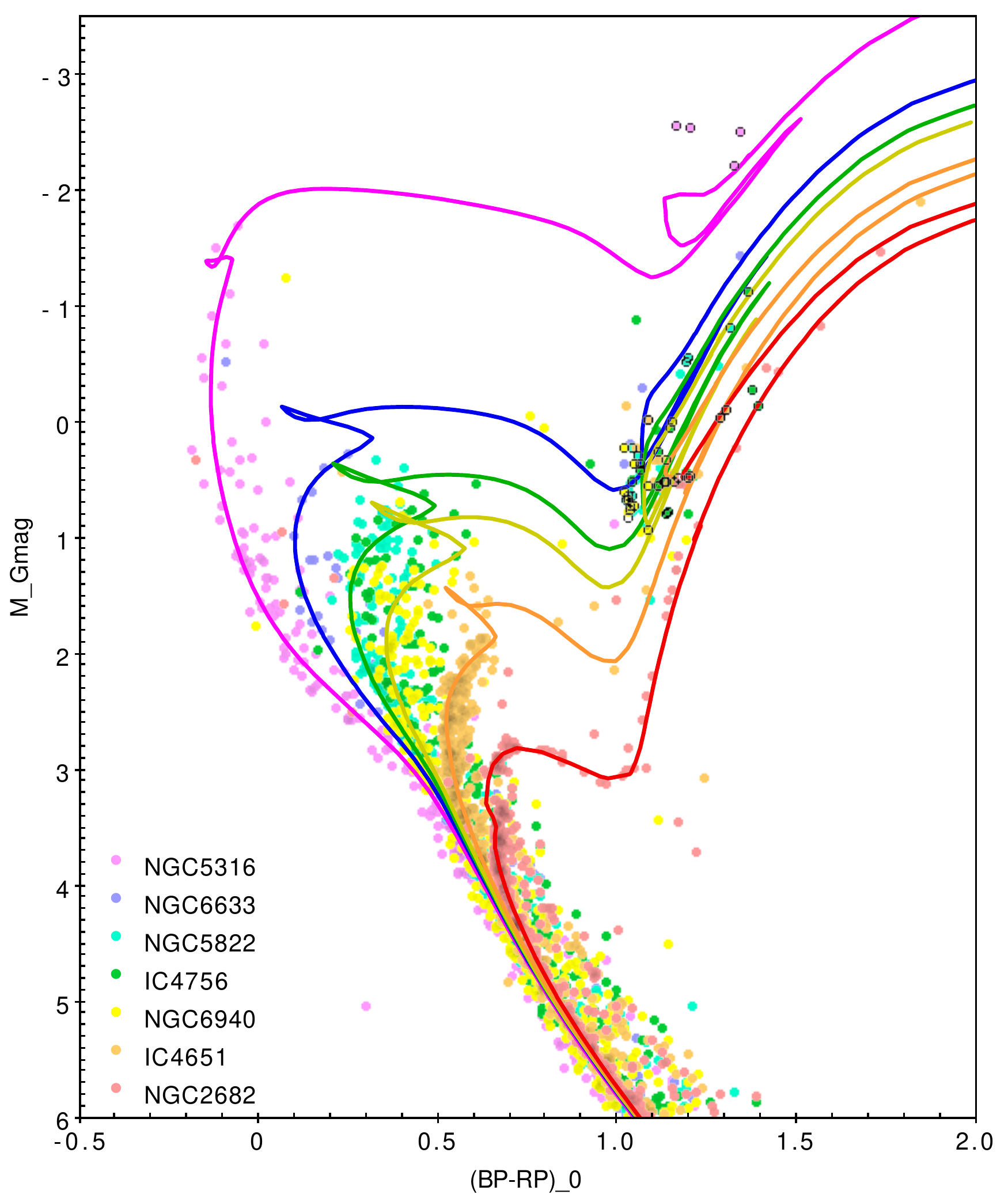}
\caption{Gaia DR2 Hertzsprung-Russell diagram for all open clusters analyzed in this work. Giant stars with FEROS spectra are marked by black circles. PARSEC isochrones \citep{Marigo+17} with the parameters presented in Table \ref{table_openclusters} are overplotted. The original Gaia DR2 data come from \citet{Cantat-Gaudin+18b}. Only stars with membership probability $\geq 0.50$ are presented.}
\label{fig_CMD_isotfit}
\end{figure}

Figure \ref{fig_CMD_isotfit} presents the Hertzsprung-Russell (HR) diagram of the seven clusters in our sample as well as the position of the 50 giant stars with FEROS spectra. This HR diagram was built using the Gaia DR2 photometry \citep{gaia18a} and \citet{Cantat-Gaudin+18b} catalogs, which include cluster membership determination. The isochrone fits obtained by \citet{gaia18b} and \citet{bossini19} %(Table \ref{table_openclusters}) 
were used to convert the observational magnitudes to the absolute ones. Table \ref{table_openclusters} shows the physical parameters of our cluster sample and the number of stars analyzed in each cluster. We directly used these cluster ages to develop the analysis presented in this work.
We adopted conservative age uncertainties of 20\% since the estimates from \citet{bossini19} are only the formal ones, which are clearly underestimated in the most cases. Furthermore, it seems that only such level of age uncertainty can explain the differences with the results found by \cite{cantat-gaudin20} analysing the same Gaia DR2 data. These last authors do not provide individual uncertainties but informed that the log(age) uncertainty ranges from 0.15 to 0.25 for young clusters and from 0.1 to 0.2 for old clusters, which are in agreement with our conservative assumption.
Table\,\ref{stasamp} shows the 50 cluster giants analyzed in this work.

\begin{center}
\begin{table*}
\footnotesize
\caption{Physical parameters of the seven open clusters analyzed in this work. [Fe$/$H] is the spectroscopic metallicity used for the isochrone fitting in each work, the last column presents the number of giant stars with FEROS spectra.}
\label{table_openclusters}
\begin{tabular}{l|cccccccc} \hline \hline
Cluster & RA (J2000) & Dec (J2000) & Age (Gyr) & [Fe$/$H] & distance (kpc) & E(B-V) & Reference & $N_{\rm{spectra}}$\\ \hline \hline
NGC\,2682 & 132.846 & $+11.814$ & 3.64$^{+0.02}_{-0.02}$ & 0.00 & 0.88 & 0.04 & \citet{bossini19} & 3 \\
 & & & 4.3 & 0.00$^*$ & 0.89 & 0.02 & \citet{cantat-gaudin20} & \\
NGC\,5316 & 208.516 & $-61.883$ & 0.153$^{+0.002}_{-0.004}$ & 0.13 & 1.44 & 0.25 & \citet{bossini19} & 4 \\
 & & & 0.17 & 0.00$^*$ & 1.45 & 0.30 & \cite{cantat-gaudin20} & \\
NGC\,5822 & 226.051 & $-54.366$ & 0.891$^{+0.000}_{-0.002}$ & 0.08 & 0.81 & 0.11 & \citet{bossini19} & 9 \\
 & & & 0.91 & 0.00$^*$ & 0.85 & 0.13 & \cite{cantat-gaudin20} & \\
IC\,4651  & 261.212	& $-49.917$ & 2.0 & 0.12 & 0.95 & 0.04 & \citet{gaia18b} & 7 \\
 & & & 1.7 & 0.00$^*$ & 0.98 & 0.14 & \citet{cantat-gaudin20} & \\
NGC\,6633 & 276.845 & $+06.615$ & 0.773$^{+0.055}_{-0.011}$ & $-0.08$ & 0.37 & 0.15 & \citet{bossini19} & 5 \\
 & & & 0.69 & 0.00$^*$ & 0.42 & 0.10 & \cite{cantat-gaudin20} & \\
IC\,4756  & 279.649 & $+05.435$ & 0.971$^{+0.031}_{-0.011}$  & 0.00 & 0.48 & 0.13 & \citet{bossini19} & 11 \\
 & & & 1.3 & 0.00$^*$ & 0.51 & 0.09 & \cite{cantat-gaudin20} & \\
NGC\,6940 & 308.626 & $+28.278$ & 1.02$^{+0.07}_{-0.07}$ & 0.15 & 1.03 & 0.15 &\citet{bossini19} & 11 \\
 & & & 1.3 & 0.00$^*$ & 1.10 & 0.13 & \cite{cantat-gaudin20} & \\
\hline \hline
\end{tabular}
\end{table*}
\end{center}

\begin{center}
\begin{table*}
%\centering
\caption{Sample of stars studied in this work. Full version of this table is available as supplementary material with the online version of this paper.}
\begin{tabular}{l|ccccc} \hline \hline
Star             &    Gaia\,DR2\,ID       &     RA         &        DEC         &   V    &  (B$-$V)  \\  \hline \hline
IC\,4756$-$12    &   4284658038773033728  & 18\,35\,47.43  &  $+$05\,20\,17.11  &  9.54  &  1.03  \\    
IC\,4756$-$14    &   4284662849136453376  & 18\,35\,58.46  &  $+$05\,24\,59.99  &  8.86  &  0.86  \\     
IC\,4756$-$28    &   4283901746580386816  & 18\,36\,33.23  &  $+$05\,12\,42.78  &  9.01  &  1.32  \\     
IC\,4756$-$38    &   4283997575895463680  & 18\,37\,05.21  &  $+$05\,17\,31.62  &  9.83  &  1.10  \\     
IC\,4756$-$42    &   4284806438475643776  & 18\,37\,20.77  &  $+$05\,53\,43.11  &  9.46  &  0.97  \\    
IC\,4756$-$44    &   4283983552796484864  & 18\,37\,30.30  &  $+$05\,12\,15.72  &  9.77  &  1.08  \\  
\end{tabular}
\label{stasamp}
\end{table*}
\end{center}

\section{Spectroscopic analysis}\label{spectros}

We applied the Local Thermodynamical Equilibrium (LTE) approximation to derive the abundance ratios in the cluster sample. The Stellar Atmospheric Parameters (SAP), effective temperature (T$_{\rm{eff}}$), surface gravity ($\log\,g$), microturbulence velocity ($\xi$) and metallicity [Fe$/$H] are prerequisite to find the chemical abundances. Specifically, T$_{\rm{eff}}$ were obtained from excitation equilibrium, $\log$\,g from ionization equilibrium between Fe\,I and Fe\,II lines, $\xi$ from zero slope between Fe\,I abundances and reduced equivalent width and [Fe$/$H] were established from the final value fixed by the ionization equilibrium. To avoid the dependency on the $\log\,gf$-values we have employed the line-by-line (lbl) method relative to the Juno solar spectrum.

All abundances were obtained via Equivalent Width (EWs) measurements, with the exception of Ba and Eu that were computed via spectral synthesis technique. To measure the EQW we used the Gaussian fit of splot$/$IRAF\footnote{IRAF is developed by NOAO and available on http://iraf.noao.edu/}, and to obtain the stellar parameters and chemical abundances we employed the python-code qoyllur-quipu\footnote{this code was developed by \citep{ram14} and it is available on https://github.com/astroChasqui/q2}. This code uses the Kurucz atmospheric grids\footnote{These grids are available on http://kurucz.harvard.edu/} and spectral code MOOG\footnote{MOOG is available in its last version on https://www.as.utexas.edu/~chris/moog.html}. The list of the Fe-lines were taken from \citet{lambert96}~$\&$~\citet{castro97} and for the other elements from several sources such as it is shown in the Table\,\ref{ic4756}. 

\begin{center}
%%\footnotesize
\begin{table*}
\caption{Electronic parameters and equivalent widths of the absorption lines in IC\,4756. The full version of this table, as well as the tables from the other clusters, are available as supplementary material with the online version of this paper.}
\begin{tabular}{cccccccccccccccc} \hline \hline
Wavelength & Species & $\chi$ & log\,$gf$ &  E12    & E14    & E28    &  E38   &  E42   & E44     &  E52   & E101   & E109   & E125   &  E164   & Juno \\  \hline \hline
5711.09    &  12     & 4.34   &  -1.73   &  124.7  & 136.6  & 133.7  &  117   &  113.9 &  112.1  &  143.3 & 120.1  & 124.2  &  120.4 &  121.8  & 101.4 \\
6318.71    &  12     & 5.11   &  -1.94   &   58.4  &  67.4  &  66.9  &  53.6  &  55.5  &   57.7  &  71.8  &  57.1  &  55.3  &  ---   &   61.2  & 41.3  \\
6319.24    &  12     & 5.11   &  -2.17   &   36.7  &  45.6  &  47.2  &  32.2  &  33.3  &   31.1  &  46.3  &  29.2  &  36.5  &  ---   &   36.7  & 28.8  \\
7387.69    &  12     & 5.75   &  -1.25   &   70.2  &  77.0  &  76.7  &  74.1  &  67.4  &   67.9  &  85.9  &  72.3  &  81.7  &  73.1  &   69.3  & 72.7  \\
7691.55    &  12     & 5.75   &  -1.00   &   71.7  &  ---   &  ---   &  ---   &  ---   &   ---   &  ---   &  ---   &  ---   &  ---   &   ---   & 56.2  \\
8712.69    &  12     & 5.93   &  -1.31   &   52.7  &  ---   &  ---   &  ---   &  ---   &   ---   &  ---   &  ---   &  ---   &  ---   &   ---   & 52.8  \\
\end{tabular}
\label{ic4756}
\end{table*}
\end{center}

We have computed [Mg$/$Fe], [Al$/$Fe], [Sr$/$Fe], [Y$/$Fe], [Zr$/$Fe], [Ba$/$Fe], [La$/$Fe], [Ce$/$Fe], [Nd$/$Fe], [Eu$/$Fe], [Y$/$Mg], [Y$/$Al], [Y$/$Si], [Y$/$Ca], and [Y$/$Ti]. This process was made for six open clusters of our sample. For IC\,4651, we have taken SAP, [Y$/$Fe], [Zr$/$Fe], [La$/$Fe], [Ce$/$Fe], and [Nd$/$Fe] from \citet{KSR17} to derive the other abundance ratios and chemical clocks. These authors used FEROS spectra and the same methodology employed by us. For our entire sample, the abundance ratios of Mg, Al, Sr, and Zr were obtained from the neutral state whereas for the elements Y, Ba, La, Ce, Nd, and Eu the ratios were derived from their first-ionization state.

We have compared our results with those from the works of \citet{lu15, luck18} in which the author has developed an homogeneous analysis of high-resolution spectra corresponding to giant and dwarf stars with ages until 8\,Gyr and 12\,Gyr, respectively. An important detail is that this author did not derive the spectroscopic age indicators, therefore we have used his results to derived [Y$/$Mg], [Y$/$Al], [Y$/$Si], [Y$/$Ca], and [Y$/$Ti]. The giant sample of \citet{lu15} will be also useful because offer a opportunity to compare the giant stars within two scenarios, the field and open clusters. The inclusion of the field dwarfs studied by the same author will allow to test if the spectroscopic clocks also work in dwarfs such as it was reported by other authors, if work, this would indicate that the spectroscopic analysis carry out by \citet{lu15, luck18} is consistent. From these papers, we have directly taken the stellar ages whereas for carry out our analysis, the cluster non-members were used as well as the stars within the metallicity interval of [-0.30, 0.30]\,dex. It is also import to add that we have selected only the field stars whose chemical abundances of YII, MgI, AlI, SiI, CaII, TiI, SrI, BaII and EuII were calculated from more than two absorption lines (with the exception of SrI and EuII). As well as we have also taken the field stars with ages derived from at least two isochrones.

For comparison, we have used the open cluster samples of literature whose ages and abundances were obtained in homogeneous form. We have used the open clusters also studied by \citet{lu15}: Me\,25, NGC\,2287, NGC\,2632, IC\,2391, NGC\,3532, Me\,111, NGC\,6475 and NGC\,6705. As well as the Reddy's sample  \citep{R12, R13, R15, R19} and the work of \citet{BC2015} who studied dwarf and giant stars within several open clusters.

\subsection{Solar abundances}

Our solar atmospheric parameters (5777$\pm$24\,K; 4.45$\pm$0.05\,dex; 7.48$\pm$0.02\,dex; 0.94$\pm$0.09\,km\,s$^{-1}$) were used to find the abundances of the chemical species studied in this work. In general terms, the solar chemical abundances of Juno spectrum are similar to the reported in literature (Table\,\ref{solabun}). However, some Juno abundances present differences $>$0.10\,dex, specially in relation to \citet{asplund09} and the giant stars in the open cluster M67 studied in Gaia-ESO iDR5.

For example, although the our determinations of AlI are similar to Sun\,iDR5 and \citet{gre07}, when it is compared to \citet{asplund09} we see that there is a difference of +0.12\,dex. In the same way, the abundance of SiI has +0.12\,dex in relation to the Giants of M67\,iDR5 and the CaI presents a difference of +0.17\,dex in relation to M67\,iDR5. Other elements with important differences are the ZrI which presents -0.12\,dex and -0.11\,dex between iRD5 (Sun and M67) and Juno respectively. Finally, chemical abundances of LaII and EuII have a difference of -0.17\,dex when they are compared to M67\,iDR5.

These differences can affect in direct way the comparisons between our values and literature. More detailed comparisons, star by star in relation to open clusters and field stars of literature will be made in section\,\ref{abundances_from_lit}. The implications of these differences in our analyses of the spectroscopic clocks will be discussed in section\,\ref{53}.

\begin{center}
\begin{table}
\footnotesize
\caption{Juno abundances compared with Gaia-ESO Sun\,iDR5, G\,2007\citep{gre07}, A\,2009\citep{asplund09} and giants M67\,iDR5.}
\label{solabun}
\begin{tabular}{l|ccccc} \\ \hline \hline
Spec  &      Juno       &    Sun\,iDR5    &      G\,2007     &     A\,2009     &   M67\,iDR5    \\ \hline \hline
FeI   &  7.48$\pm$0.06  &  7.49$\pm$0.03  &   7.50$\pm$0.05  &  7.50$\pm$0.04  &  7.45$\pm$0.01  \\
MgI   &  7.55$\pm$0.07  &  7.51$\pm$0.07  &   7.53$\pm$0.09  &  7.60$\pm$0.04  &  7.51$\pm$0.05  \\
AlI   &  6.33$\pm$0.03  &  6.34$\pm$0.04  &   6.37$\pm$0.06  &  6.45$\pm$0.03  &  6.41$\pm$0.04  \\
SiI   &  7.43$\pm$0.07  &  7.48$\pm$0.06  &   7.51$\pm$0.04  &  7.51$\pm$0.03  &  7.55$\pm$0.06  \\
CaI   &  6.27$\pm$0.06  &  6.31$\pm$0.12  &   6.31$\pm$0.04  &  6.34$\pm$0.04  &  6.44$\pm$0.10  \\
TiI   &  4.92$\pm$0.07  &  4.90$\pm$0.08  &   4.90$\pm$0.06  &  4.95$\pm$0.05  &  4.90$\pm$0.09  \\
SrI   &  2.85           &        ---      &   2.97$\pm$0.07  &  2.87$\pm$0.07  &        ---      \\
YII   &  2.23$\pm$0.05  &  2.19$\pm$0.12  &   2.21$\pm$0.02  &  2.21$\pm$0.05  &  2.14$\pm$0.09  \\
BaII  &  2.20$\pm$0.07  &  2.17$\pm$0.06  &   2.13$\pm$0.05  &  2.18$\pm$0.09  &  2.07$\pm$0.07  \\ 
ZrI   &  2.65$\pm$0.09  &  2.53$\pm$0.13  &   2.60$\pm$0.02  &  2.58$\pm$0.04  &  2.54$\pm$0.05  \\ 
LaII  &  1.17$\pm$0.02  &       ---       &   1.17$\pm$0.07  &  1.10$\pm$0.04  &  1.00$\pm$0.12  \\ 
CeII  &  1.60$\pm$0.06  &  1.70$\pm$0.11  &   1.58$\pm$0.07  &  1.58$\pm$0.04  &  1.71$\pm$0.01  \\
NdII  &  1.51$\pm$0.08  &        ---      &   1.50$\pm$0.06  &  1.42$\pm$0.04  &       ---       \\ 
EuII  &  0.59           &  0.52$\pm$0.06  &   0.51$\pm$0.08  &  0.52$\pm$0.04  &  0.42$\pm$0.04  \\   \hline
\end{tabular}
\end{table}
\end{center}

\subsection{Stellar parameters and abundances}

The SAP obtained in this work are in the intervals: T$_{eff}$\,=\,[4347, 5335]K $\log$\,g\,=\,[1.36, 3.38]\,dex; $\xi$\,=\,[1.45, 2.82]km\,s$^{-1}$; and [Fe$/$H]\,=\,[-0.15, 0.17]\,dex. Therefore the obtained SAP, shown in Table\,\ref{sampletab}, indicate that our sample correspond to cool RGB stars with solar metallicity. The abundances of the chemical species derived here are shown in Tables\,\ref{cab} and \ref{abu4651}.

\begin{figure}
\includegraphics[scale=0.45]{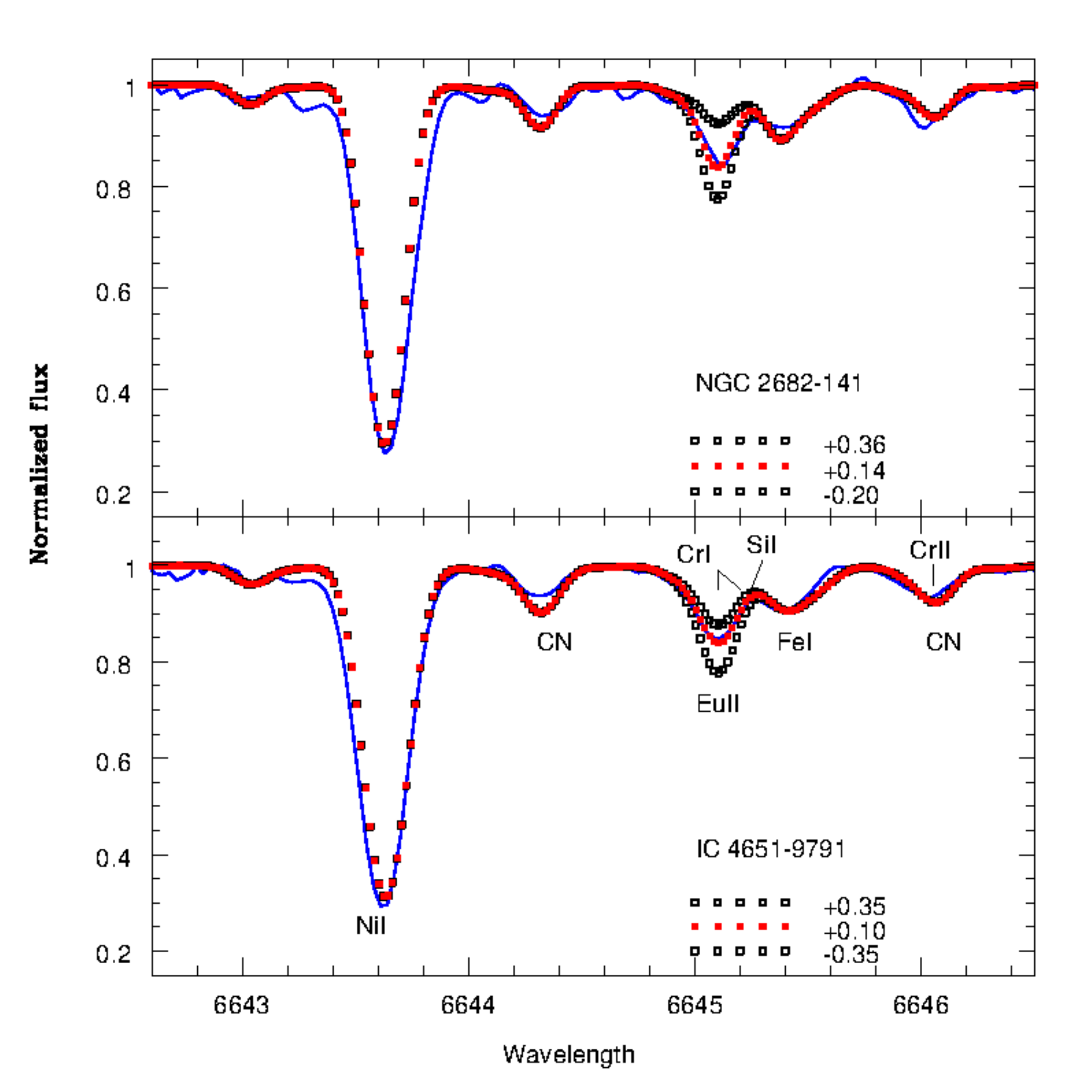}
\caption{Spectral synthesis to the Eu\,II line at $\lambda$\,6645\AA\,. Black dotted lines are the synthetic spectra whereas red dotted lines represent the final solutions. Other absorption lines are identified.}
\label{spectEu}
\end{figure}

The absorption line 4607\,\AA,~whose profile is well-defined in our spectra (Figure\,\ref{specreg}), was used to obtain the Sr abundances. To measure the Ba abundances, we used the absorption lines 5853\,\AA, 6141\,\AA~and~6497\,\AA~and taken into account the hyperfine structure (HFS) corrections of \citet{mac88}. In the case of Eu, the abundance calculations were more complicated because the line 6645\,\AA~always presented a blending with Cr\,I and Si\,I. Following the process explained in \citet{bensby05}, these absorption lines and the CN bands were included in the line list around 6645\,\AA~and therefore reliable Eu abundances were obtained in our sample. Figure\,\ref{spectEu} shows the spectral synthesis of Eu\,II in the stars NGC\,2682$-$141 and IC\,4651$-$9791. For the first time, we provide chemical abundance determinations of Sr, Ba, and Eu for some stars in IC\,4651, NGC\,5822, and NGC\,2682. The EuII line 4129\AA~never presented a well-defined profile and therefore abundances from this line were not obtained. The HFS corrections of \citet{mucciarelli08} were used to derive the Eu abundances. 

\begin{figure*}
\includegraphics[scale=0.4]{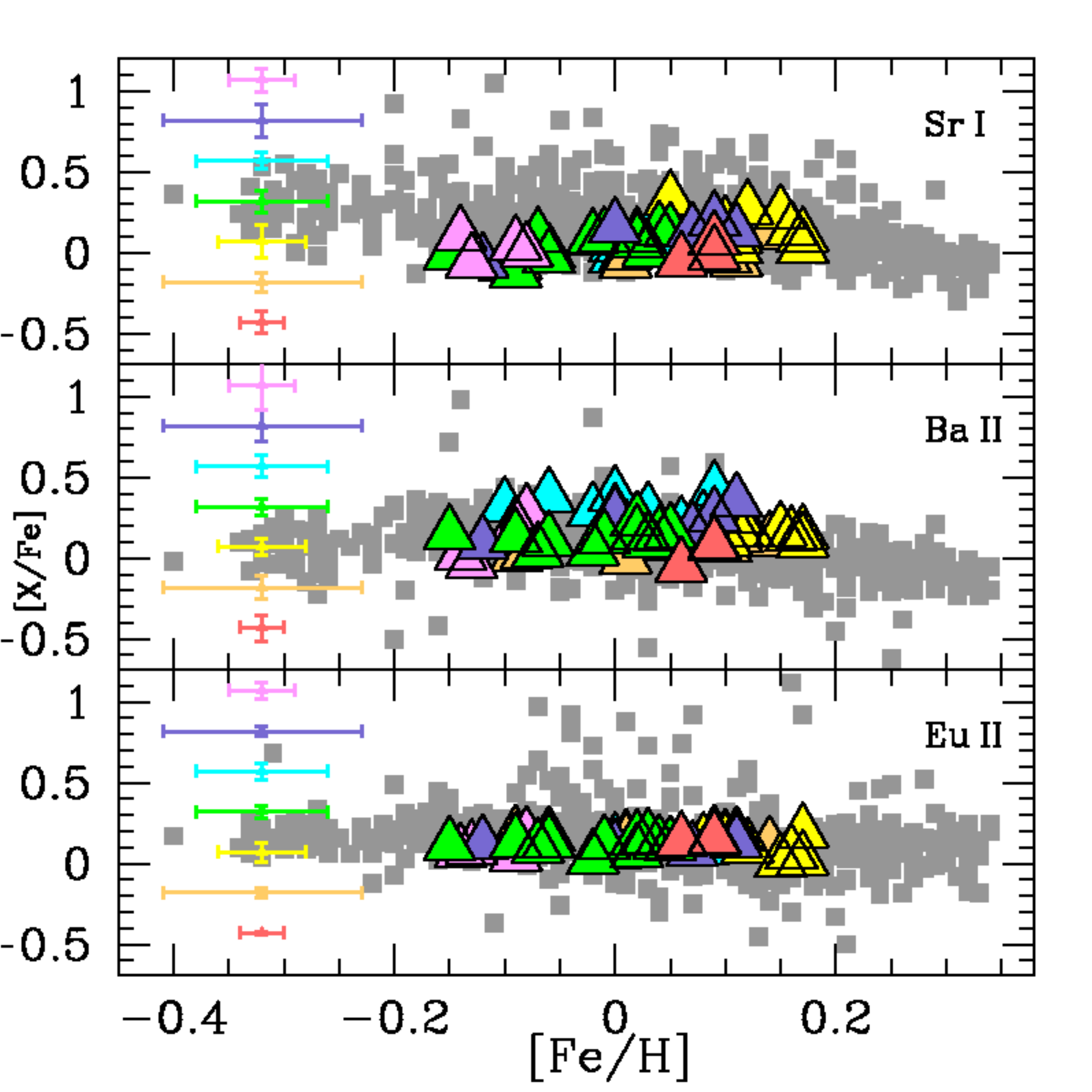}
\includegraphics[scale=0.4]{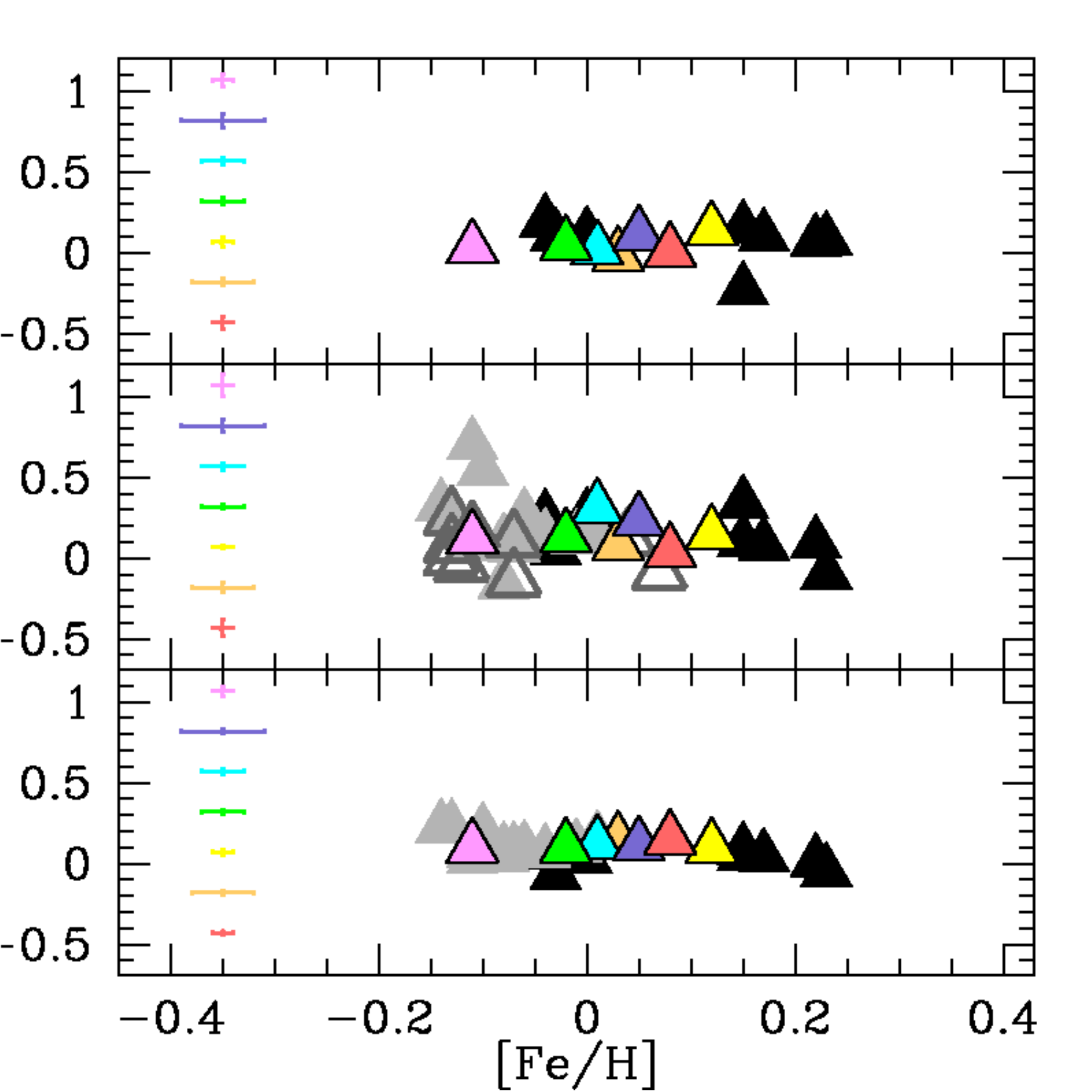}
\caption{Left: Chemical abundances of Sr, Ba $\&$ Eu star by star in IC\,4756 (green triangles), IC\,4651 (light orange triangles), NGC\,5316 (magenta clear triangles), NGC\,6633 ( blue triangles), NGC\,6940 (yellow triangles), NGC\,5822 (cyan triangles) and NGC\,2682 (light red triangles) compared with field giants of \citet{lu15} (gray squares). Error bars represent the standard deviation. Right: comparisons with open cluster samples studied from giant stars: \citet{lu15} (black triangles); \citet{R12, R13, R15, R19} (grey triangles) and \citet{BC2015} (open triangles). Error bars represent the standard deviation of the mean.}
\label{SrBaEu}
\end{figure*}

Figure\,\ref{SrBaEu} shows the chemical abundances of Sr, Ba and Eu obtained in this work. We compared our abundances with the field giants and also with \textit{Gaia} and Hi-Res open clusters. As it can be observed, there is similarity between field and open cluster giants. Therefore our abundances of Sr, Ba and Eu are consistent to the previously found in the Galactic disk. 

\begin{center}
\begin{table*}
\small
\caption{Stellar atmospheric parameters calculated in this work. The line numbers of Fe\,I and Fe\,II are represented by $\#_{1}$ and $\#_{2}$. Whereas $\sigma$ and $\sigma_{n}$ represent the standard deviation and standard error of the mean, respectively. Full version of this table is available as supplementary material.}
\label{sampletab}
\begin{tabular}{l|cccc} \hline \hline
Star            & T$_{eff}\,\pm\,\sigma$ & $\log\,g\pm\,\sigma$ & $\xi\,\pm\,\sigma$ & [Fe$/$H]$\pm\,\sigma$ ($\#_{1}$,\,$\#_{2}$)  \\
                &       K         &      dex        &  km\,s$^{-1}$   &             dex                  \\ \hline\hline
IC\,4756$-$12   &   5124$\pm$21   &  2.96$\pm$0.09  &  1.61$\pm$0.06  &    -0.02$\pm$0.08 (76,\,14)      \\
IC\,4756$-$14   &   4765$\pm$52   &  2.56$\pm$0.12  &  1.74$\pm$0.08  &    -0.06$\pm$0.11 (76,\,14)      \\
IC\,4756$-$28   &   4689$\pm$34   &  2.34$\pm$0.15  &  1.76$\pm$0.07  &    -0.15$\pm$0.11 (75,\,14)      \\
IC\,4756$-$38   &   5224$\pm$27   &  3.20$\pm$0.08  &  1.66$\pm$0.06  &     0.02$\pm$0.07 (76,\,14)      \\
IC\,4756$-$42   &   5228$\pm$29   &  3.16$\pm$0.08  &  1.61$\pm$0.06  &     0.03$\pm$0.08 (76,\,14)      \\
IC\,4756$-$44   &   5177$\pm$28   &  3.17$\pm$0.08  &  1.58$\pm$0.07  &     0.02$\pm$0.08 (77,\,13)      \\
\end{tabular}
\end{table*}
\end{center}

\begin{center}
\begin{table*}
\footnotesize
\caption{Abundance ratios [x$/$Fe] for the elements calculated in this work. These ratios were calculate from [X$/$Fe]\,=\,[X$/$H]\,-\,[Fe$/$H] where [X$/$H] is obtained from [X$/$H]\,=\,$\log\,(N_{\rm X}/N_{\rm H})_{\star}\,-\,\log\,(N_{\rm X}/N_{\rm H})_{\odot}$. Full version of this table is available as supplementary material.}
\label{cab}
\begin{tabular}{l|ccccccccccccc} \hline \hline
star	     &    [Mg$/$Fe] & [Al$/$Fe] & [Si$/$Fe]  &   [Ca$/$Fe]  &  [Ti$/$Fe]  & [Sr$/$Fe] & [Y$/$Fe]  &  [Zr$/$Fe]  &  [Ba$/$Fe]  & [La$/$Fe]  &  [Ce$/$Fe]  & [Nd$/$Fe]  &  [Eu$/$Fe]	\\ \hline \hline
IC\,4756-12  &      0.04  &  0.07   &	 0.11  &	0.13	&	0.07	&  0.09   &	0.15	&	-0.09	&	0.05	&	0.21	&	0.18	&	0.20	&	0.03	\\
IC\,4756-14  &      0.04  &  0.11   &	 0.18  &	0.09	&  -0.03	& -0.03   &	0.16	&	-0.11	&	0.11	&	0.38	&	0.24	&	0.25	&	0.13	\\
IC\,4756-28  &      0.08  &  0.28   &	 0.20  &	0.10	&	0.00	& -0.01   &	0.09	&	-0.13	&	0.15	&	0.33	&	0.14	&	0.01	&	0.11	\\
IC\,4756-38  &     -0.04  &  0.04   &	 0.07  &	0.06	&	0.02	&  0.07   &	0.14	&	 0.01	&	0.15	&	0.29	&	0.15	&	0.16	&	0.15	\\
IC\,4756-42  &     -0.07  &  0.02   &	 0.09  &	0.06	&  -0.02	&  0.01   &	0.08	&	-0.02	&	0.10	&	0.18	&	0.11	&	0.03	&	0.16	\\
IC\,4756-44  &     -0.07  &  0.05   &	 0.06  &	0.08	&  -0.01	&  0.10   &	0.16	&	-0.07	&	0.23	&	0.63	&	0.20	&	0.14	&	0.07	\\  
\end{tabular}
\end{table*}
\end{center}

\begin{center}
\begin{table*}
\footnotesize
\caption{Abundance ratios for Mg, Al, Si, Ca, Ti, Sr, Ba, and Eu in the cluster IC\,4651.}
\label{abu4651}
\begin{tabular}{l|cccccccc} \hline \hline
Star            &  [Mg/Fe]  &  [Al/Fe]  &  [Si/Fe]  &  [Ca/Fe]  &  [Ti/Fe]  &  [Sr/Fe]  &  [Ba/Fe]  &  [Eu/Fe]	\\ \hline
IC\,4651-6333   &	0.07	&	0.18    &   0.08    &   0.06    &    0.03	&	-0.06	&	0.00	&	0.15	\\
IC\,4651-7646   &	0.05	&	0.18    &   0.07    &  -0.05    &   -0.07	&	-0.03	&	0.07	&	0.15	\\
IC\,4651-8540   &	0.00	&	0.13    &  -0.01    &  -0.17    &   -0.17	&	-0.07	&	0.01	&	0.18	\\
IC\,4651-9025   &	0.05	&	0.22    &   0.00    &  -0.10    &   -0.11	&	-0.05	&	0.13	&	0.17	\\
IC\,4651-9122   &	0.04	&	0.19    &   0.25    &   0.11    &    0.12	&	-0.01	&	0.18	&	0.13	\\
IC\,4651-9791   &	0.11	&	0.24    &   0.26    &   0.10    &    0.08	&	-0.03	&	0.03	&	0.10	\\
IC\,4651-14527  &	0.09	&	0.13    &   0.17    &   0.10    &    0.09	&	0.12	&	0.11	&	0.11	\\ \hline
Mean$\pm\sigma_{n}$  & 0.06$\pm$0.01  & 0.18$\pm$0.02  & 0.12$\pm$0.04  &  0.01$\pm$0.04  &  0.00$\pm$0.04   & -0.02$\pm$0.02 & 0.08$\pm$0.03  & 0.14$\pm$0.01  \\ \hline  \hline
\end{tabular}
\end{table*}
\end{center}

\subsection{Abundance uncertainties}

In order to evaluate the effect of the stellar atmospheric parameters in the final chemical abundances, we have obtained the propagation of the uncertainty in all calculated abundances. Furthermore, we also included the line to line abundance scatter as another error source. The total error of the abundances were obtained from the quadratic sum of the previously variables mentioned. In Table\,\ref{erroe38} are shown these errors for the star IC\,4756-38. This star has errors similar to the other stars studied here. 

\begin{center}
\begin{table}
\footnotesize
\caption{Error propagation in the abundances for IC\,4756$-$38. Second to sixth column gives the error from T$_{eff}$, $\log$\,g, [Fe$/$H] and line to line scatter respectively. Last column represents the total abundance error.}
\label{erroe38}
\begin{tabular}{l|cccccc} \hline \hline
Species & $\Delta\,T_{eff}$ & $\Delta\log$\,g & $\Delta\xi$ & $\Delta$\,[Fe$/$H] & $\Delta$\,Ab & $\left(\sum\,\sigma^2 \right)^{1/2}$ \\ \hline \hline
Fe\,I    &      0.02         &        0.00     &     0.02    &    0.01    &   0.01   &  0.03    \\
Fe\,II   &      0.02         &        0.04     &     0.02    &    0.01    &   0.02   &  0.06    \\
Mg\,I    &      0.01         &        0.01     &     0.01    &    0.00    &   0.03   &  0.03    \\
Al\,I    &      0.01         &        0.01     &     0.01    &    0.00    &   0.02   &  0.02    \\
Sr\,I    &      0.03         &        0.01     &     0.04    &    0.00    &   ---    &  0.05    \\
Y\,II    &      0.00         &        0.04     &     0.03    &    0.01    &   0.02   &  0.05    \\
Zr\,I    &      0.04         &        0.01     &     0.00    &    0.01    &   0.06   &  0.07    \\
%Ba\,II  &      0.01         &        0.02     &     0.04    &    0.01    &   ---    &  0.07    \\
La\,II   &      0.00         &        0.04     &     0.01    &    0.01    &   0.03   &  0.05    \\
Ce\,II   &      0.01         &        0.04     &     0.01    &    0.01    &   0.01   &  0.04    \\
Nd\,II   &      0.00         &        0.04     &     0.02    &    0.01    &   0.05   &  0.06    \\ \hline
%Eu\,II  &      0.01         &        0.04     &     0.01    &    0.01    &   ---    &  0.04  \\ \hline 
\end{tabular}
\end{table}
\end{center}

\section{Spectroscopic age indicators} \label{53}

It is known from the stellar evolution theory that Interstellar medium (ISM) pollution from low-mass AGB stars increases the abundances of Y and Ba with time \citep{trav04,fish17}. Otherwise, considering that Mg and Al elements are mainly produced by core-collapse SNe, the final stage of high-mass stars (>8M$_{\odot}$), and that the ISM chemical enrichment occurs in short time-scales, it is reasonable to assume that [Mg$/$Fe] and [Al$/$Fe] decrease with time \citep{matt14}. Consequently, we can deduce that younger stars have higher [Y$/$Mg] and [Y$/$Al] ratios than older stars.

This behavior is found when solar twins are analyzed from high-resolution spectroscopy \citep{niss15,TM16,spin16,spin18}. According to \citet{fel17}, the [Y$/$Mg]$-$age relation only works for solar analogue stars because it depends on the [Fe$/$H]. This same behaviour was also reported by \citet{DM19} using F\,G\,K field dwarf stars. In fact, they have found that the linear functions to calculate the ages, using spectroscopic clocks, are limited by metallicity. These last results do not match those found by \citet{ambre} who, considering solar neighbourhood turn-off stars, have concluded that [Y$/$Mg] negative trend with stellar ages is independent of the metallicity.

\begin{figure*}
\includegraphics[scale=0.34]{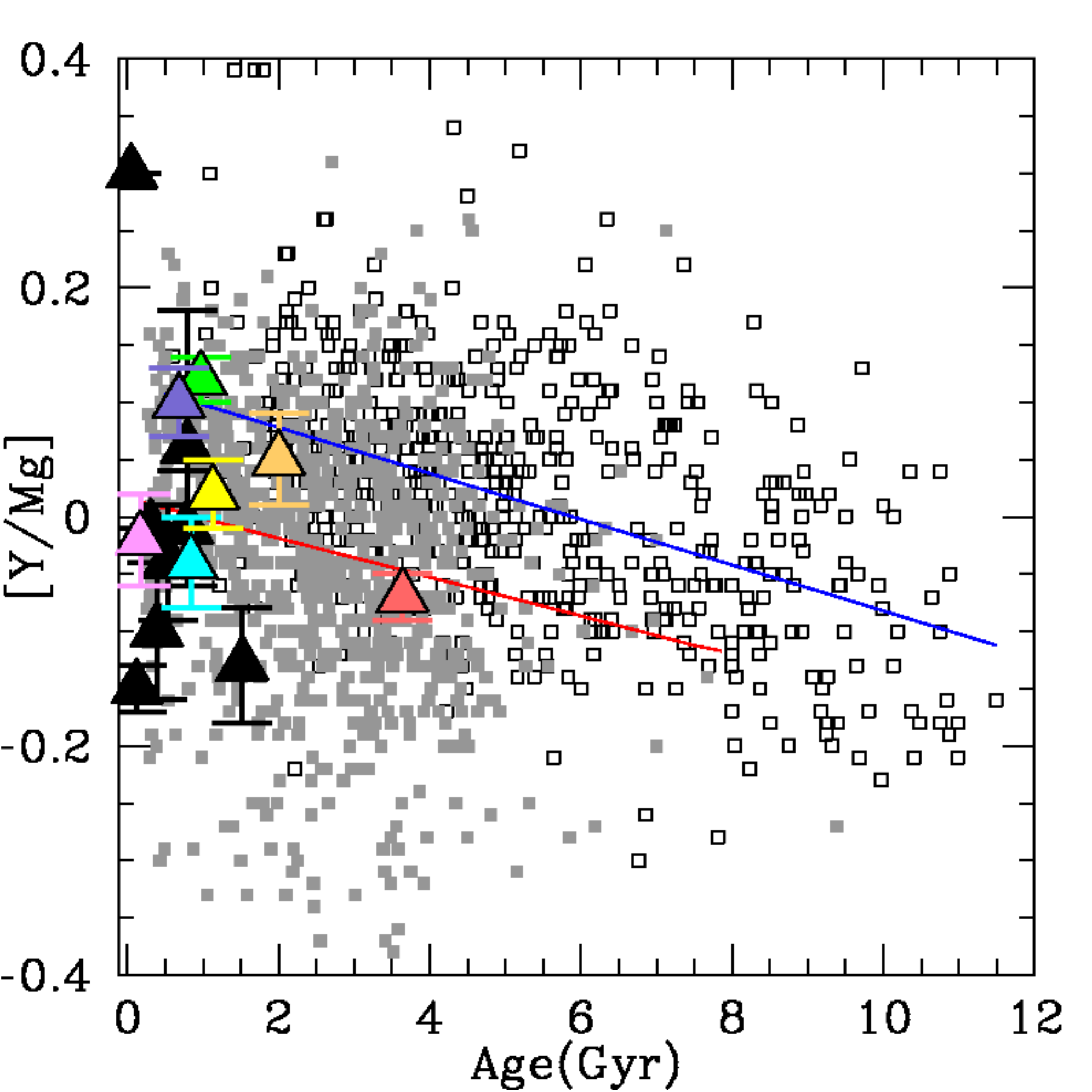}
\includegraphics[scale=0.34]{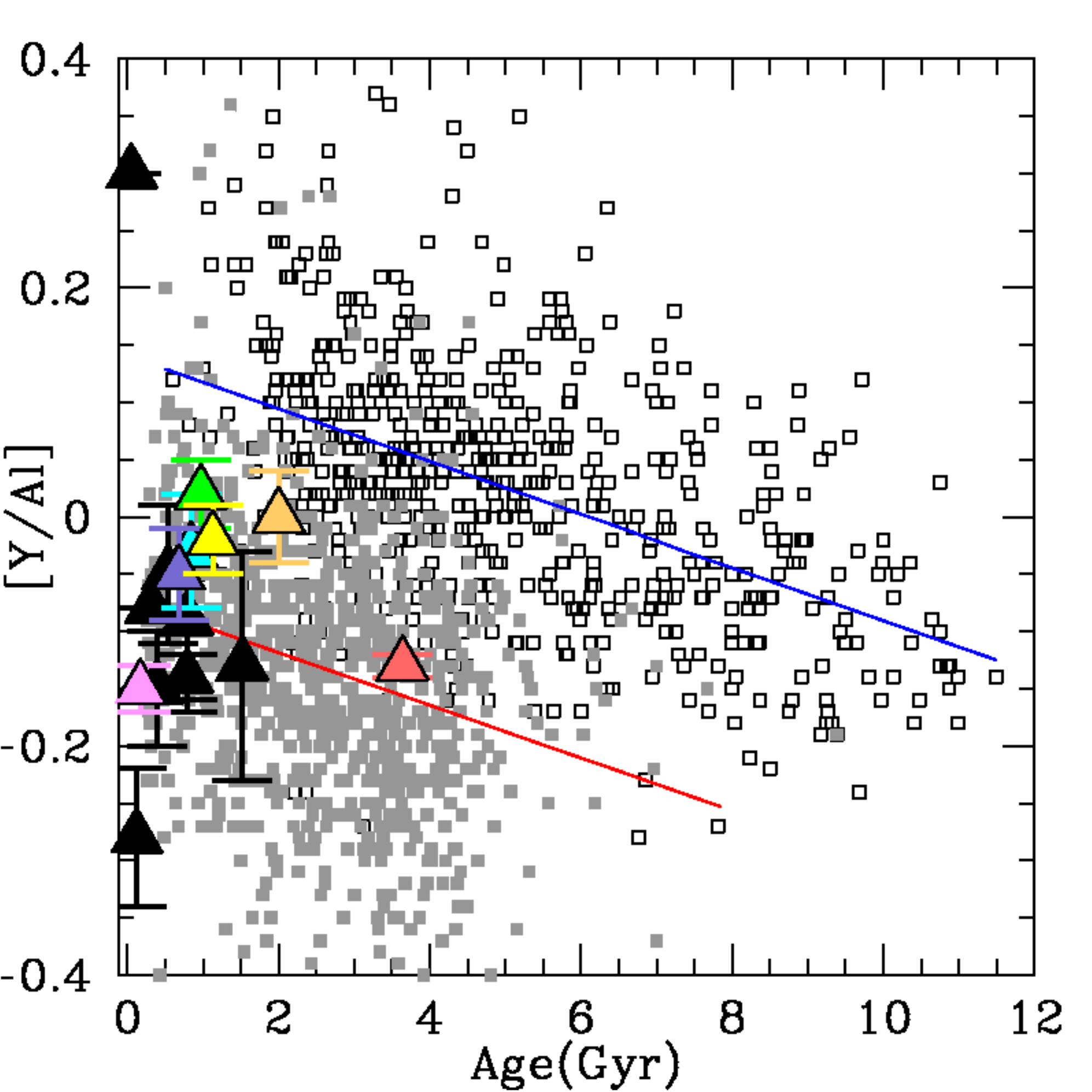}
\includegraphics[scale=0.34]{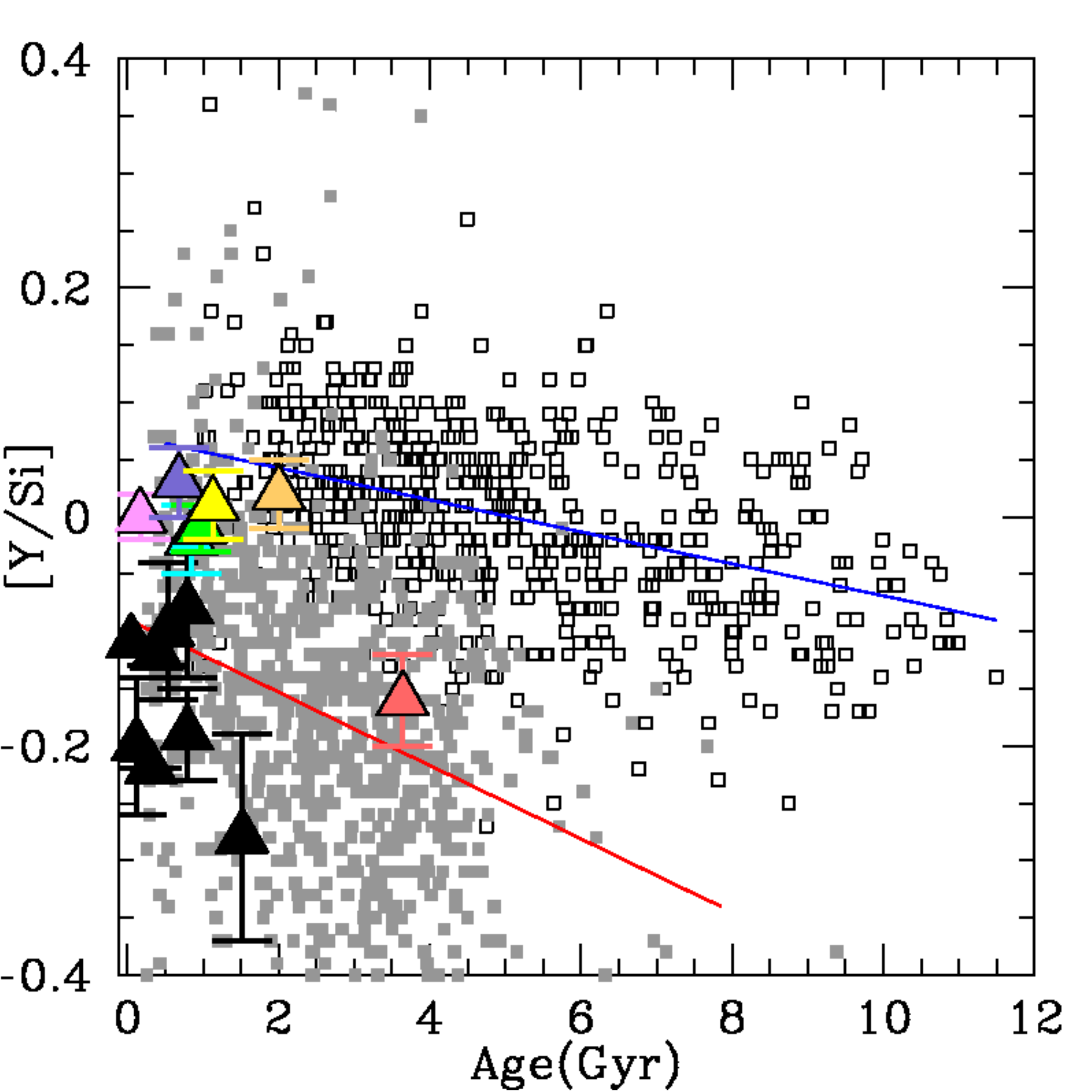}
\includegraphics[scale=0.34]{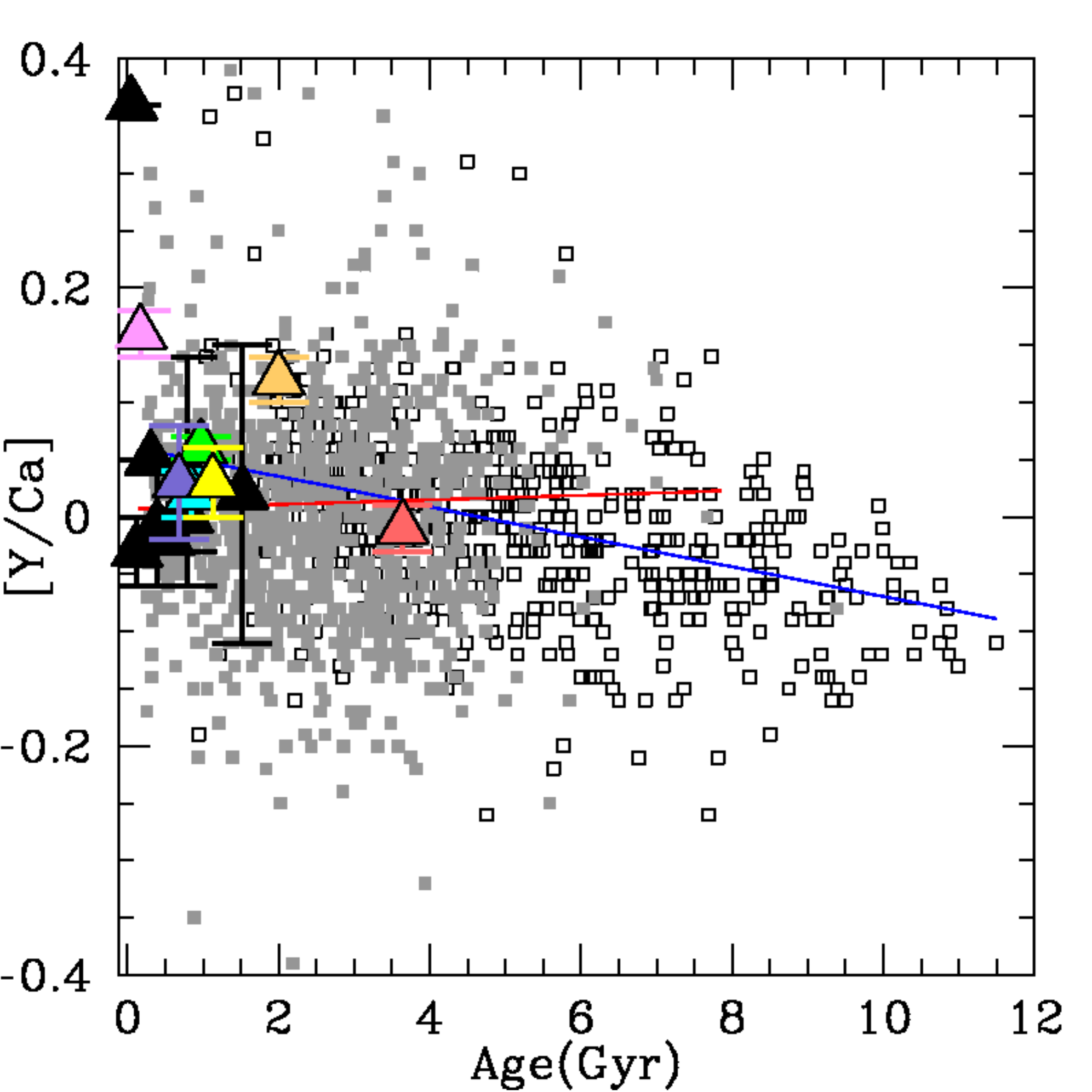}
\includegraphics[scale=0.34]{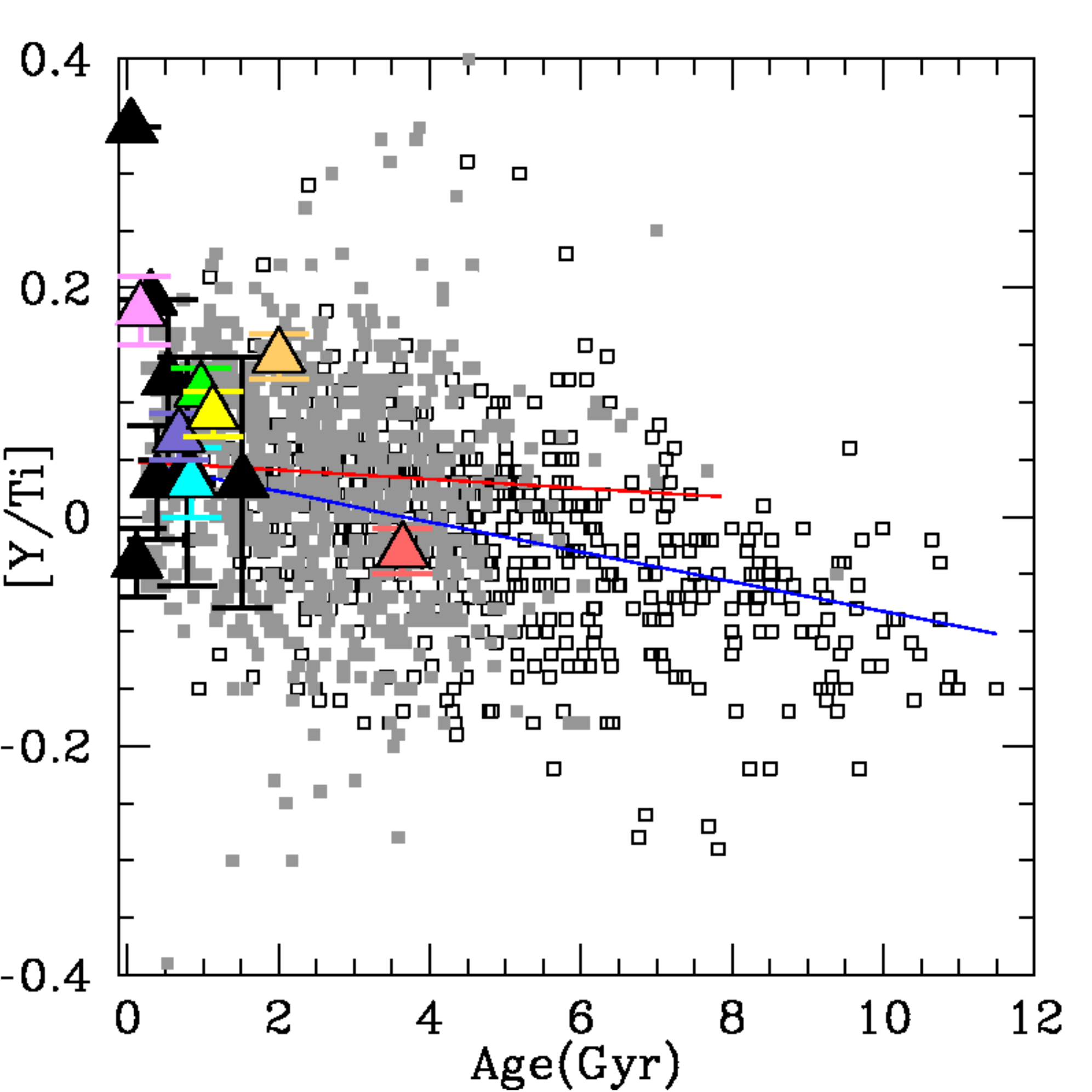}
\caption{Abundance ratios as function of age for our sample, field giants (grey squares) and open clusters (black triangles) of \citep{lu15} as well as the field dwarfs (open squares) of \citep{luck18}. The red and blue lines represent the least squares fitting for the giants and dwarfs respectively. (See Table \ref{linear_fit_luck15}).}
\label{YallOCRC}
\end{figure*}

\citet{slum17} analyzed five stars in the clusters NGC\,6811, NGC\,6819, M\,67, and NGC\,188, with Fe$/$H] close to the solar metallicity. This paper concluded that the [Y$/$Mg] clock also works for giant stars in the helium-core-burning phase (i.e., Red Clump stars). More recently, \citet{ps18} analyzed giant stars in the open clusters NGC\,2360, NGC\,3680, and NGC\,5822 (which is in common with the present work). They concluded that [Y$/$Mg] seems to be promising as age indicator but they also pointed out the necessity to have clusters with a wider age range and more studied stars per cluster. 

The first work reporting a non-universality of the spectroscopic clocks was \citet{casali}, attributing this fact to differences in Galactocentric distances or star formation histories. However, in their fig. 7, these authors found that the abundances ratios [Y$/$Mg], [Y$/$Al], [Y$/$Ti], [Y$/$Ca], and [Y$/$Si], for a sample of 19 open clusters from the Gaia-ESO survey present a high scattering in the age range <\,2.0\,Gyr. Surprisingly these authors did not examine the possibility that the origin of an apparent non-universality of the spectroscopic clocks could reside on the use of giant stars despite dwarf stars.

\subsection{Giants and dwarfs in the local region}
\label{clocks-giants-field}

Figure\,\ref{YallOCRC} shows the relations [Y$/$Mg], [Y$/$Al], [Y$/$Si], [Y$/$Ca], [Y$/$Ti]\,vs.\,age for the giant and dwarf stars of \cite{lu15, luck18}, our open clusters and also the clusters analyzed by \citet{lu15}. Further, we present the result for linear least squares regressions for the field stars. The slope ($S$) and y-axis intercept ($y$) parameters, as well as the Pearson correlation coefficients ($r$), are presented in Table \ref{linear_fit_luck15}. The uncertainties in these parameters, which correspond to 84$\%$ confidence intervals, were calculated using the bootstrapping technique. It also allowed us to estimate the $p$-values under the null hypothesis that the correlation coefficient is null. We performed linear fits using the whole samples of \cite{lu15, luck18}.

As can be seen from Table \ref{linear_fit_luck15}, for field giants the [Y$/$Mg], [Y$/$Al], and [Y$/$Si] are weakly anti-correlated with age ($r \lesssim -0.30$), besides we can assume that exist a non-null correlation with more than 99$\%$ confidence level since the $p$-value is lower than 0.01. Possibly the correlations were dilute by the scattering that was caused by the uncertainties in abundance ratios and age, but unfortunately \cite{lu15} did not provide this information. From the standard deviation of the abundance ratio values for the age range between 1 and 2\,Gyr, we estimated that the typical uncertainties range from $\sim$\,0.10\,dex ([Y$/$Mg] and [Y$/$Al]) to $\sim$\,0.10\,dex ([Y$/$Si]). Furthermore, assuming the age uncertainties are the standard deviation of the ages that were obtained by \cite{lu15} using different stellar evolutionary models, the typical relative age uncertainties would be $\sim$\,40$-$60$\%$. Concerning [Y$/$Ca] and [Y$/$Ti], the almost null $S$ and $r$ values, as well as the high $p-$values, clearly attest that these parameters do not correlate with age.

The situation for the field dwarfs is different, they present a clear anti-correlation between abundance ratios and stellar ages, with $-0.48\,\leq\,r\,\leq\, -0.36$ and $p-$values$\,\ll 0.01$. This includes [Y$/$Ca] and [Y$/$Ti], whose correlation is null for giant stars from \cite{lu15}. From comparisons with the Pearson coefficients in the dwarf stars of \citet{casali}, we could think that the works of \citet{lu15, luck18} have no high precision in their abundances and in this form producing the scattering found in the spectroscopic clocks. However, it is necessary consider that the samples of \citet{lu15, luck18} were homogeneously studied and therefore their differences in the Pearson coefficients and $p-$values may be a real fact. In other words, the giant stars could have intrinsically higher scattering than the dwarfs.

\begin{center}
\begin{table*}
%\centering
\caption{Results from the linear least square regression for the abundance ratios vs. age relations for the giant and dwarf stars presented in \citet{lu15, luck18}. The slopes ($S$) and y-axis intercept ($y$) parameters are presented, as well as the Pearson correlation coefficients ($r$) and $p-$values. The index $g$ and $d$ correspond to giants and dwarfs respectively.}
\begin{tabular}{c|cccc|cccc} \hline
Ratio         &      S$_{g}$     &       y$_{g}$      &        $r_{g}$  &        $p_{g}$            & S$_{d}$      &     y$_{d}$      &       $r_{d}$  &         $p_{d}$     \\ \hline \hline
$[$Y$/$Mg$]$  &  -0.017$\pm$0.003 &    0.016$\pm$0.010 &  -0.17$\pm$0.04 &  1.0\,$\times\,10^{-7}$   & -0.020$\pm$0.002 &  0.118$\pm$0.009 & -0.48$\pm$0.03 & 6.3\,$\times\,10^{-34}$     \\
$[$Y$/$Al$]$  &  -0.023$\pm$0.003 &   -0.072$\pm$0.010 &  -0.22$\pm$0.04 &  1.2\,$\times\,10^{-10}$  & -0.023$\pm$0.002 &  0.140$\pm$0.010 & -0.49$\pm$0.03 & 1.2\,$\times\,10^{-35}$    \\
$[$Y$/$Si$]$  & -0.032$\pm$0.004  &   -0.089$\pm$0.011 &  -0.28$\pm$0.03 &  8.4\,$\times\,10^{-17}$  & -0.014$\pm$0.001 &  0.071$\pm$0.008 & -0.38$\pm$0.03 & 3.3\,$\times\,10^{-21}$     \\
$[$Y$/$Ca$]$  &  0.002$\pm$0.003  &    0.007$\pm$0.009 &  +0.02$\pm$0.03 &         0.572             & -0.013$\pm$0.001 &  0.061$\pm$0.008 & -0.37$\pm$0.03 & 3.1\,$\times\,10^{-20}$     \\
$[$Y$/$Ti$]$  & -0.004$\pm$0.003  &    0.049$\pm$0.009 &  -0.05$\pm$0.04 &         0.139             & -0.013$\pm$0.001 &  0.048$\pm$0.008 & -0.36$\pm$0.04 & 3.0\,$\times\,10^{-19}$    \\  \hline \hline
\end{tabular}
\label{linear_fit_luck15}
\end{table*}
\end{center}

\subsection{Open cluster giants}

\begin{figure*}
\includegraphics[scale=0.2]{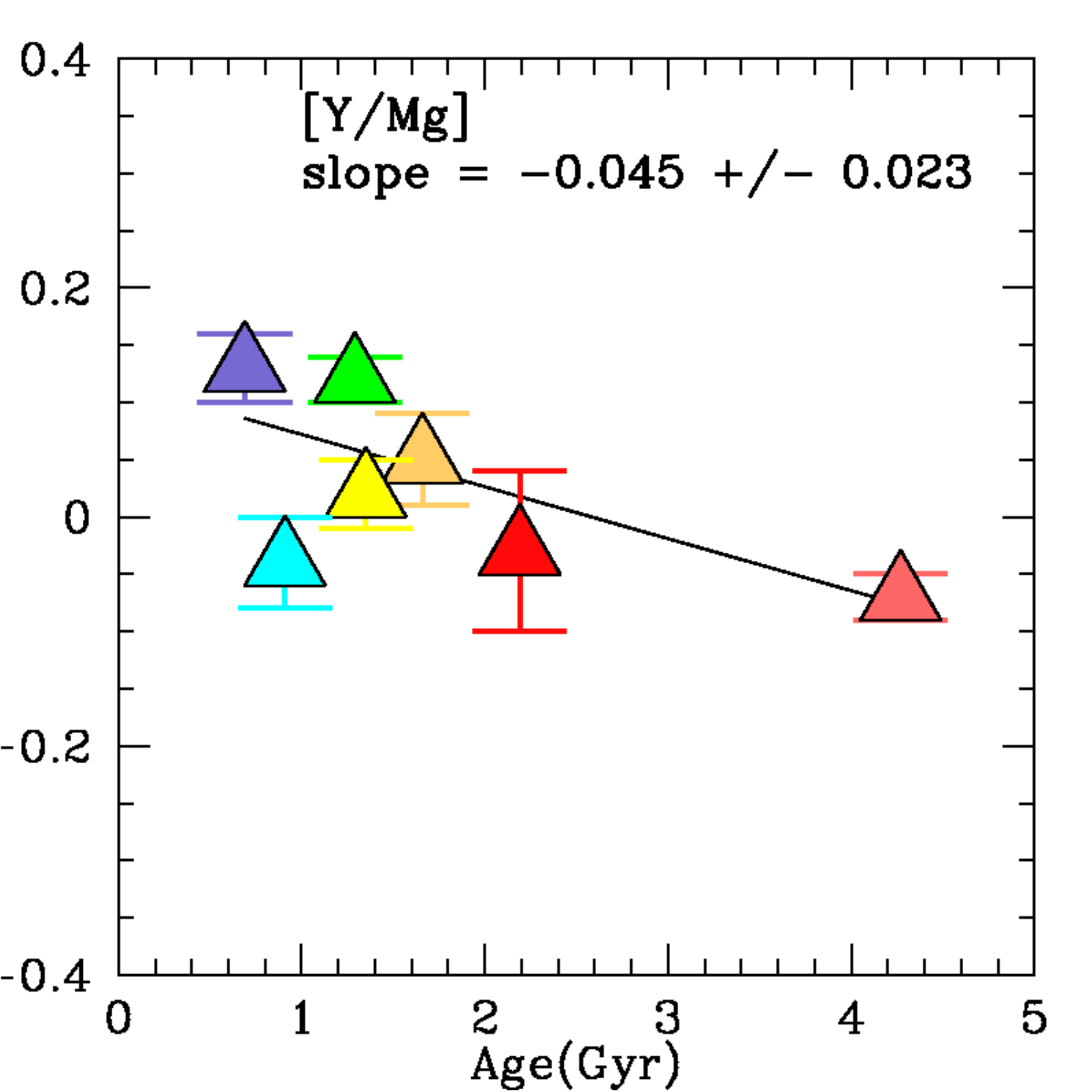}
\includegraphics[scale=0.2]{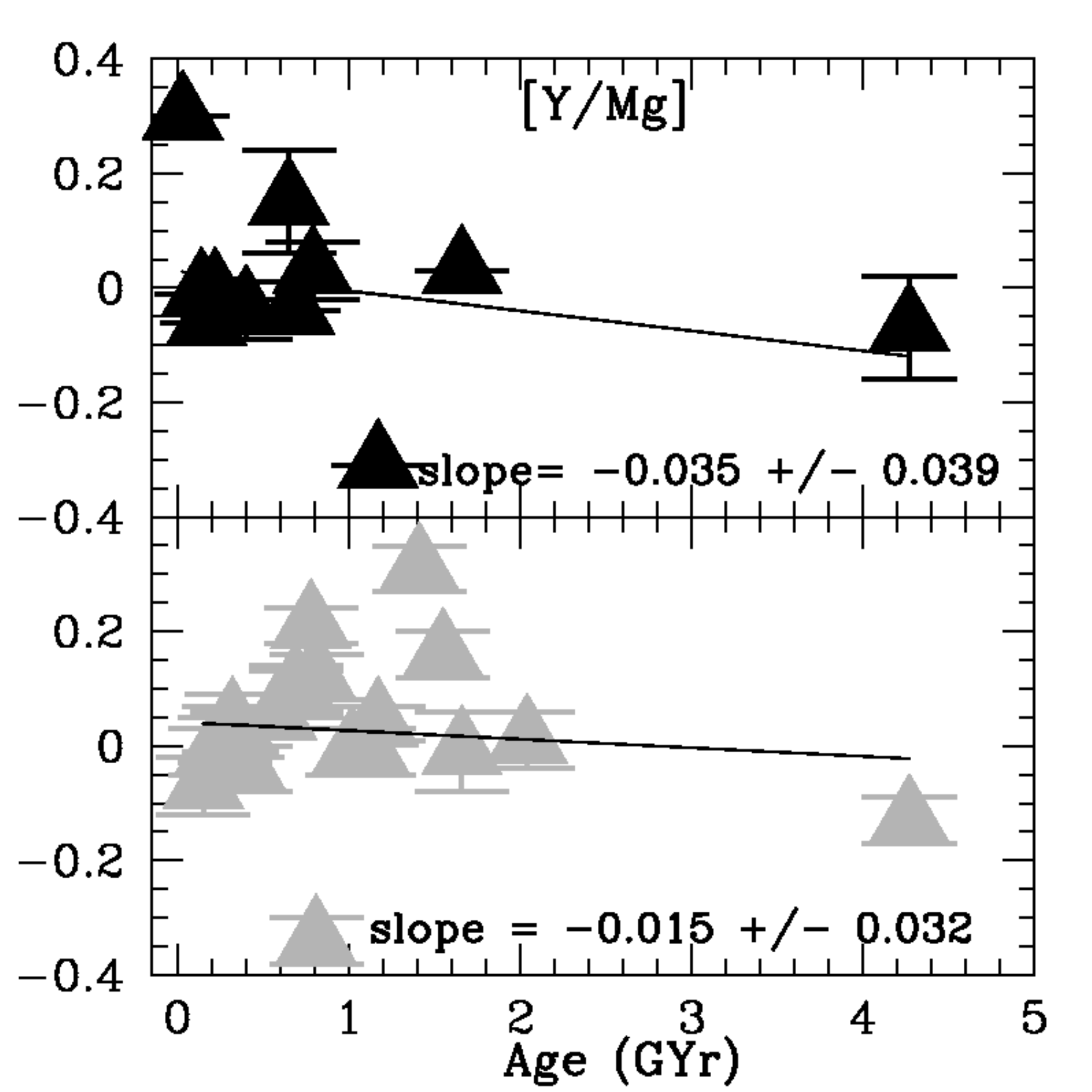}
\includegraphics[scale=0.2]{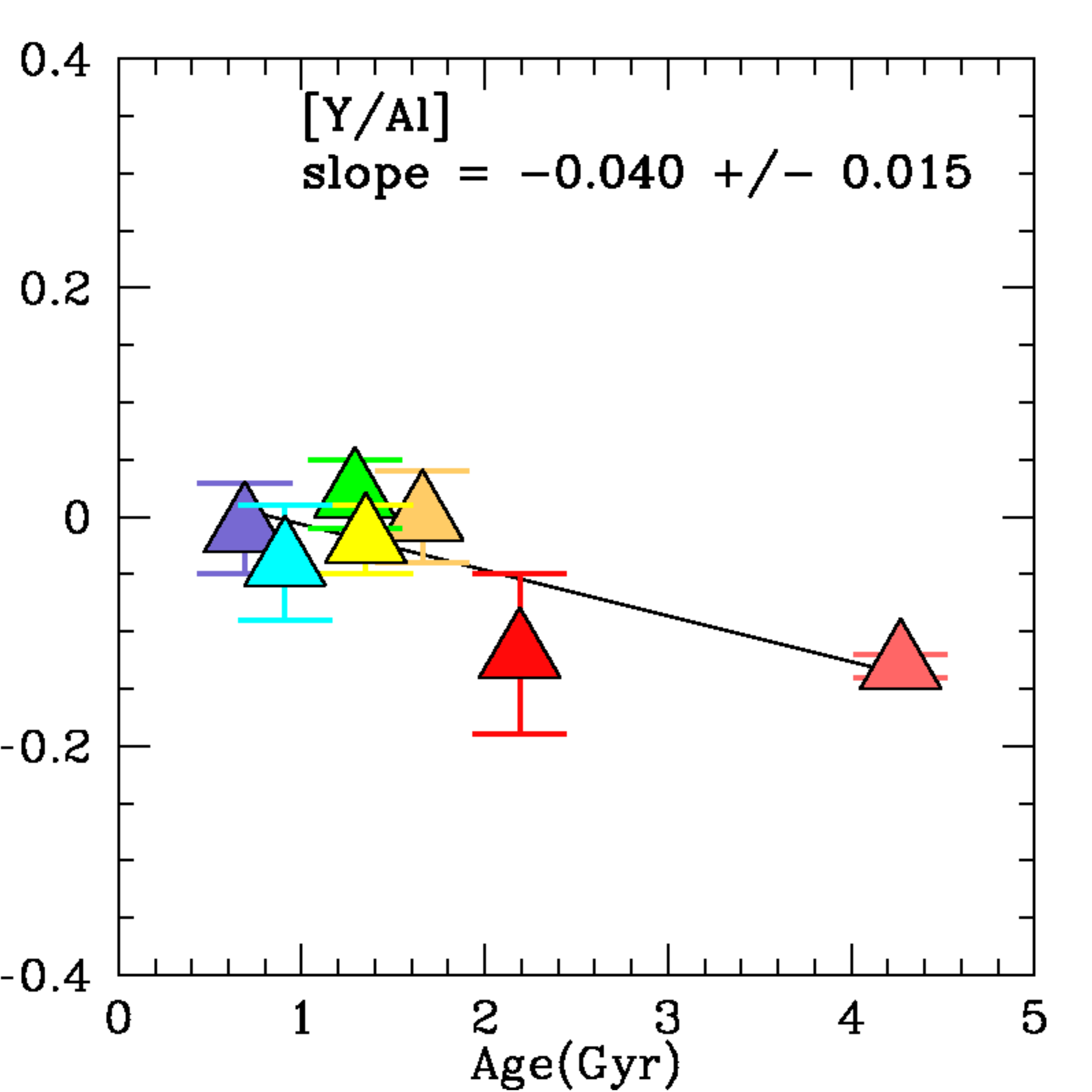}
\includegraphics[scale=0.2]{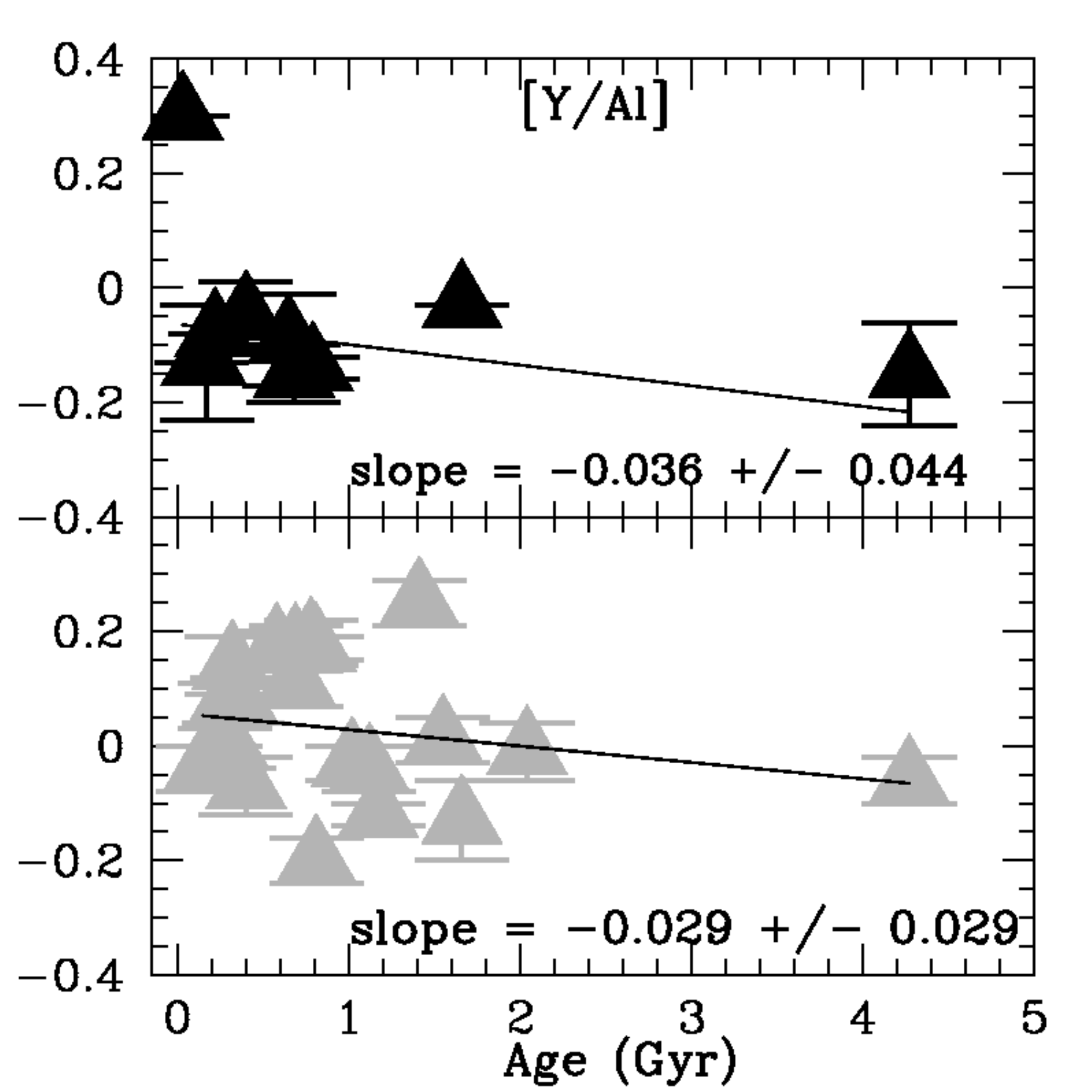}
\includegraphics[scale=0.2]{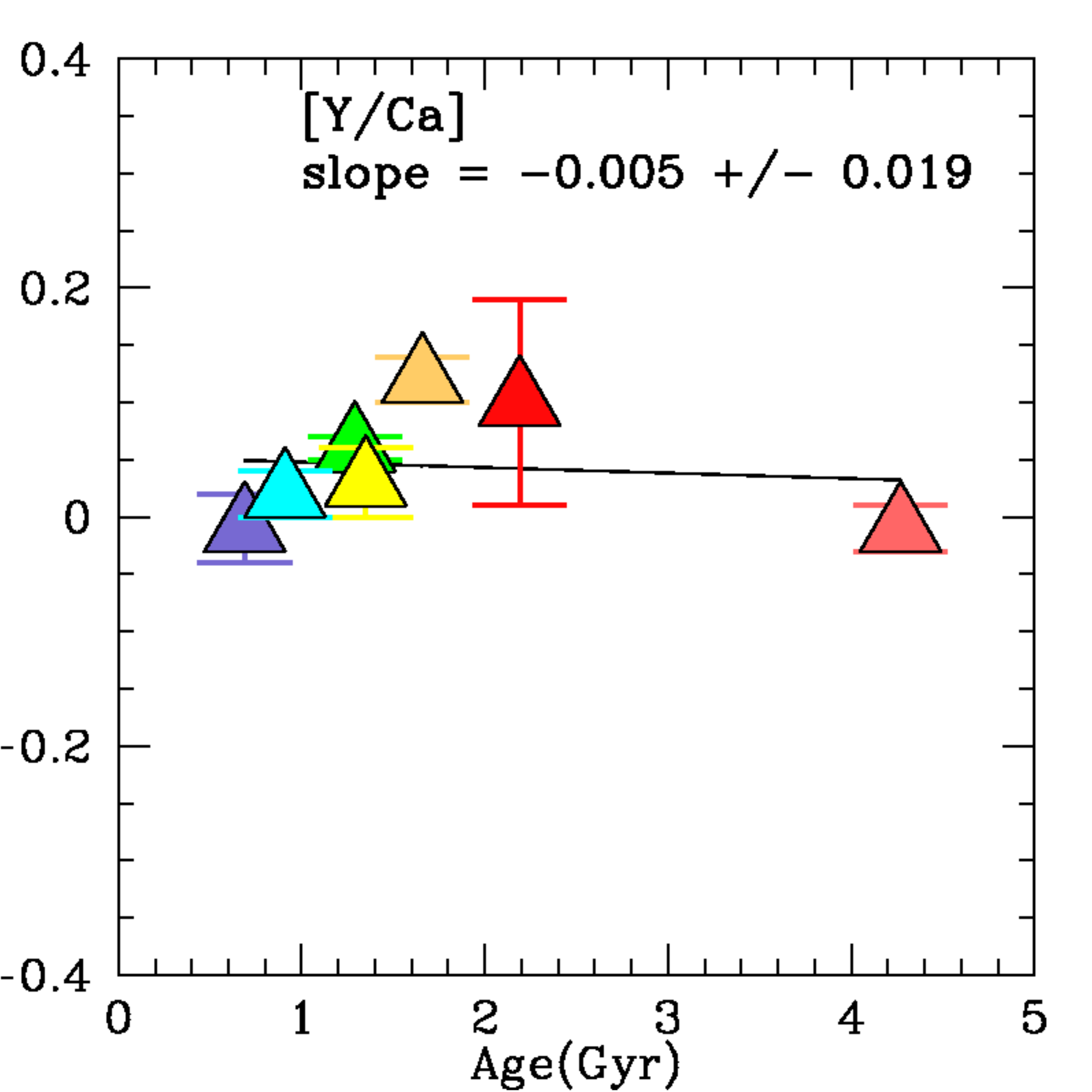}
\includegraphics[scale=0.2]{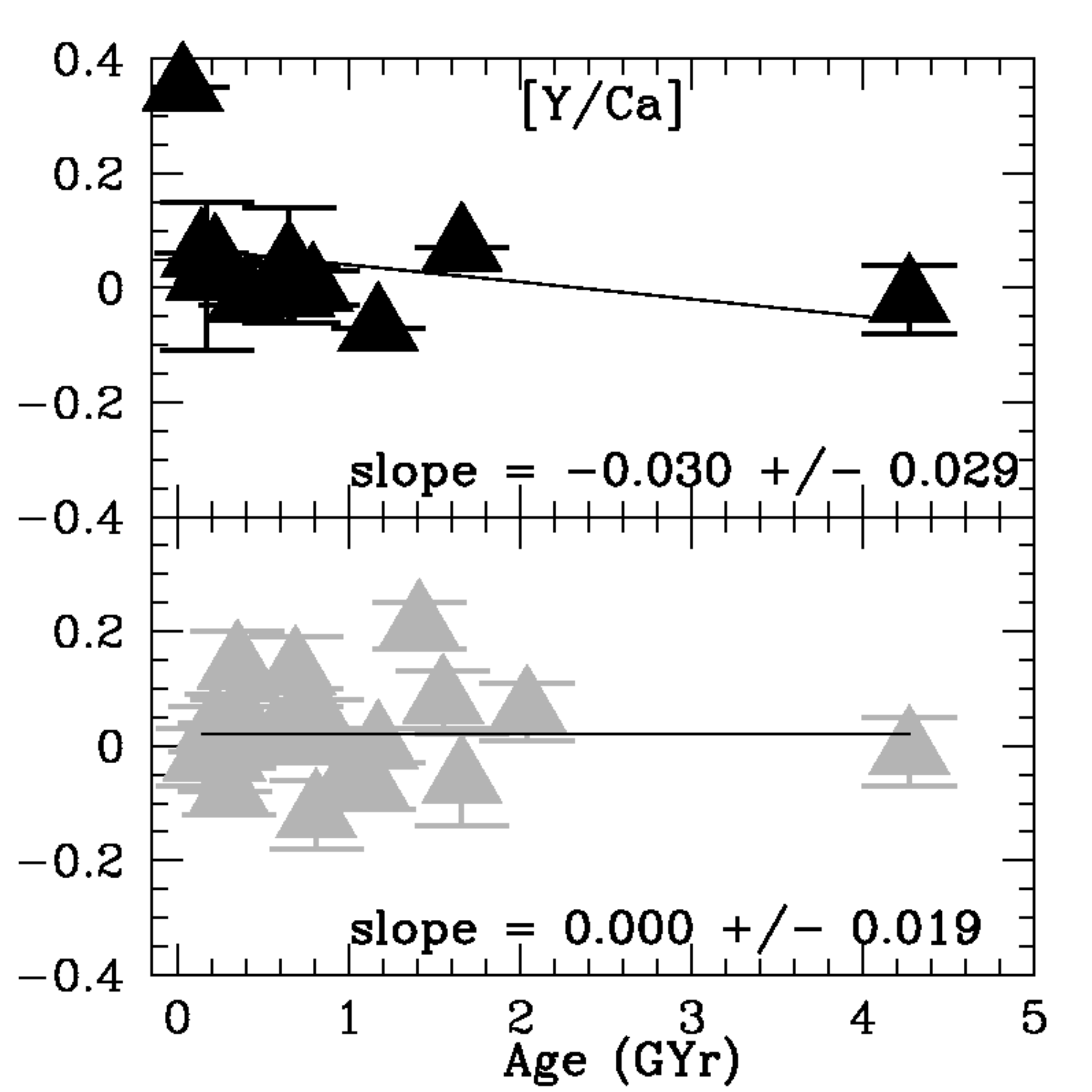}
\includegraphics[scale=0.2]{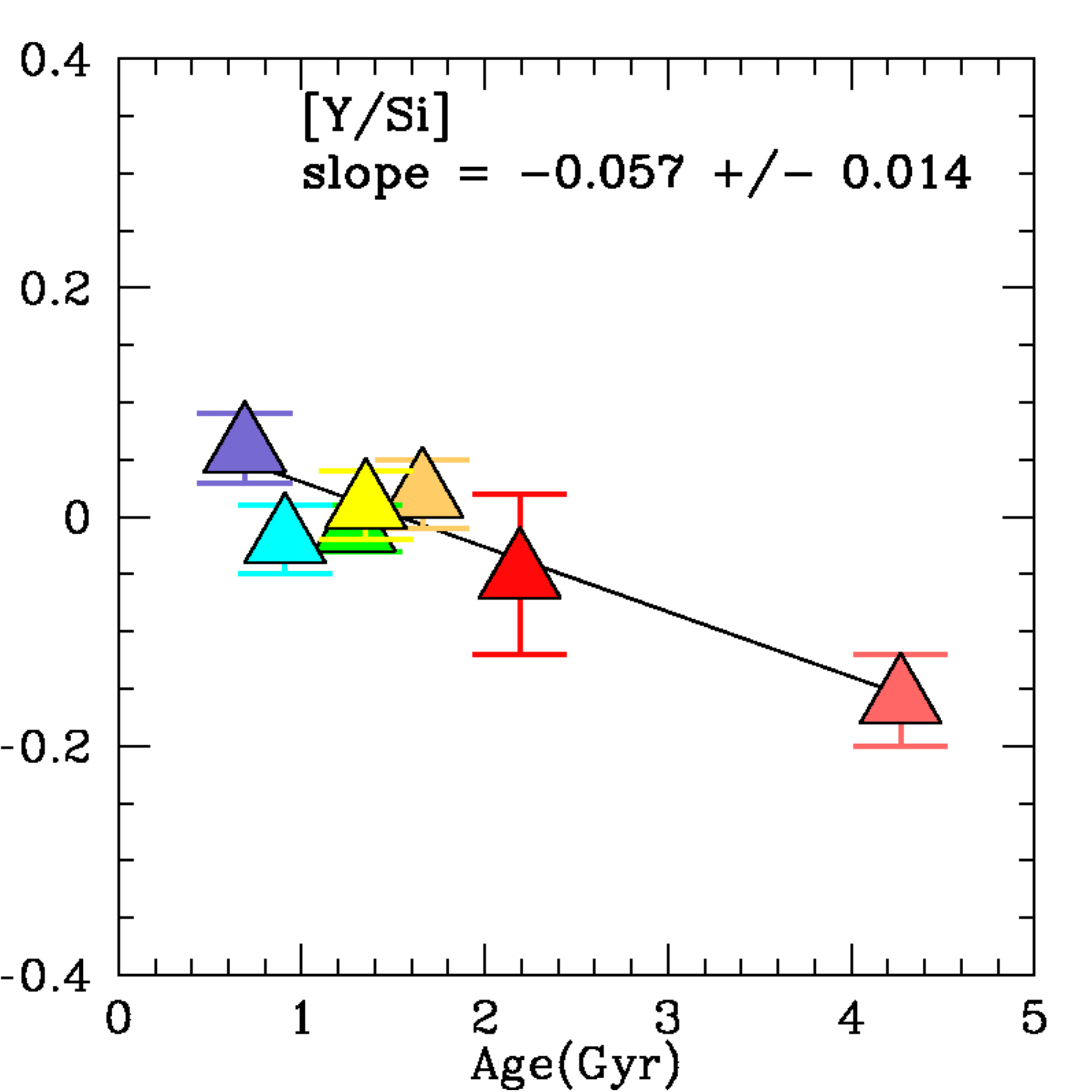}
\includegraphics[scale=0.2]{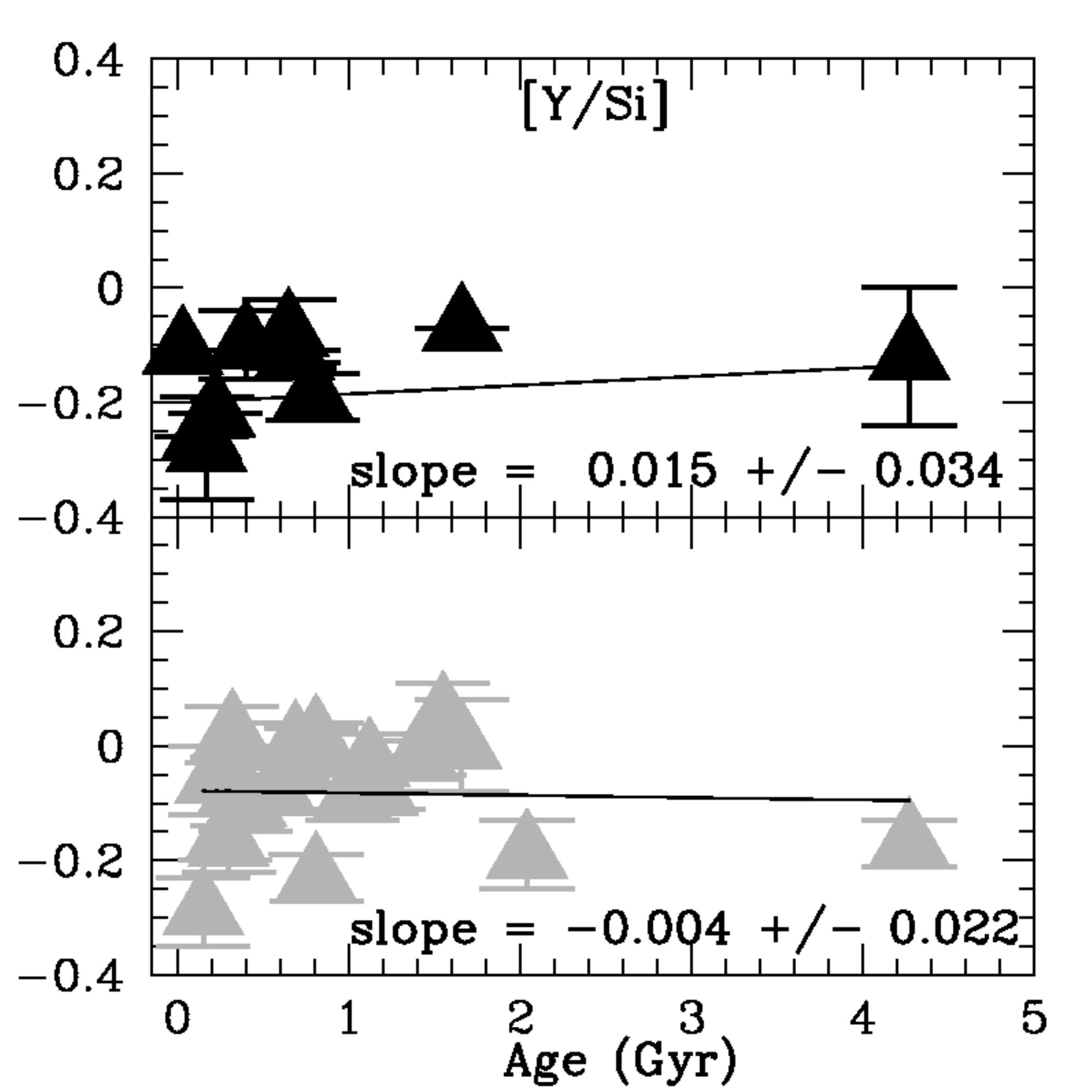}
\includegraphics[scale=0.2]{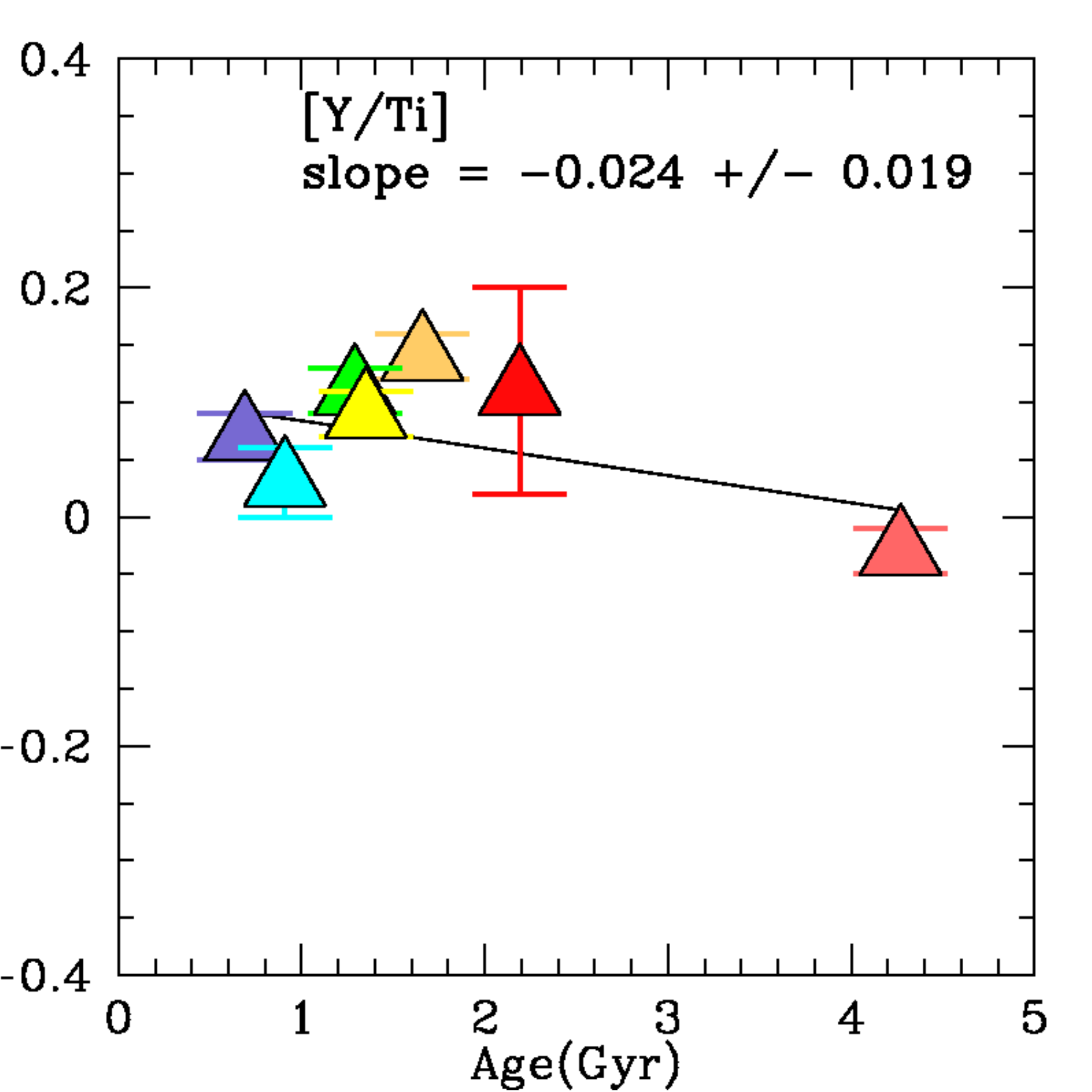}
\includegraphics[scale=0.2]{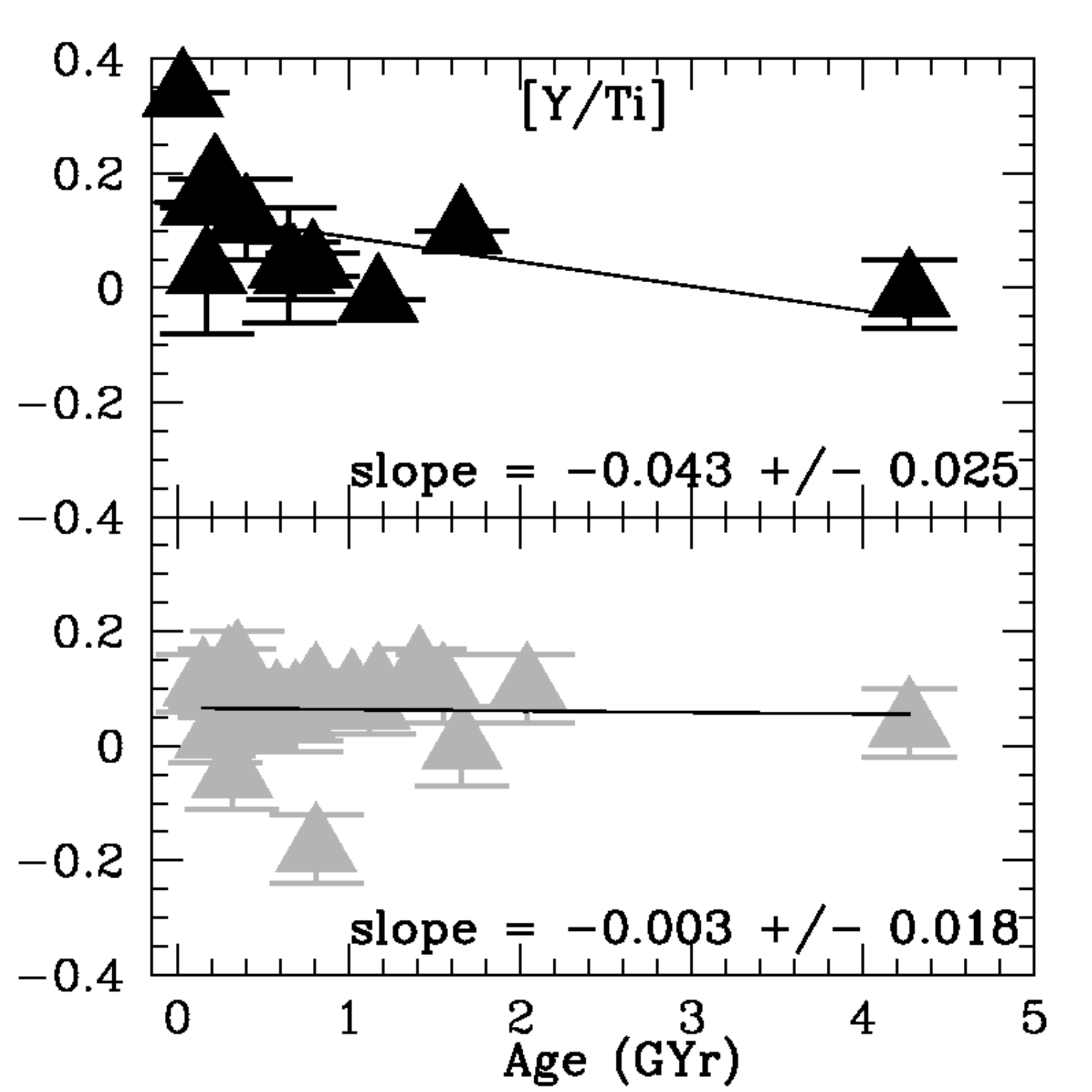}
\caption{Chemical clocks for our clusters and samples of \citet{lu15}  and \citet{R12, R13, R15, R19}. All samples have open clusters with d\,<\,1kpc. Black lines are the least-square fitting for the samples. Red triangle is the open cluster NGC\,3680 \citep{ps18}.}
\label{clockslitoc}
\end{figure*}

We have included the open cluster NGC\,3680 within our sample because \citet{ps18} used the same observational configuration and line lists than us, we have also compiled their abundances to obtain the chemical clocks, our results for the seven clusters are shown in Table\,\ref{clockabun7oc}. Figure\,\ref{clockslitoc} show a comparison between our sample and literature. In the two cases the open clusters are restricted to the local region (d\,$<$\,1\,kpc). We have found that our sample falls along a very probable tendency with the ages as it has been previously reported. This fact is specially notable for the clocks [Y$/$Mg], [Y$/$Al], [Y$/$Si] and [Y$/$Ti] whereas for the ratio [Y$/$Ca] the slope is less steeper. From our results it also possible to see that our slopes are similar to the obtained slopes for the field stars of \citet{lu15, luck18} as it can be observed in Table\,\ref{linear_fit_luck15}. In spite that we have a small sample, we seen that scatter reported by \citet{casali} and \citet{casami21} at the early ages is also present in our clusters.

To verify this early scatter, we have compiled other samples from literature. We found that the works of \citet{R12, R13, R15, R19} and \citet{BC2015} obtained chemical abundances and ages for several open clusters in homogeneous form. Despite that these authors derived abundances and ages, they did not discuss the spectroscopic clocks. We have used these results and applied the same restriction (d\,$<$\,1\,kpc) to these clusters. These derived clocks are directly compared to our results in Figure\,\ref{clockslitoc}.

From last figure we seen that the scatter seems to exist despite the restriction (d\,$<$\,1kpc). In these two samples there is also the same problem than in ours, no points between $\sim$\,2.0\,GYr and $\sim$\,4.0\,GYr and also not clusters with ages $>$4.0\,GYr. However, the few old clusters are not an exclusive problem of this work, \citet{casali} and \citet{casami21} also have few points and it is those older clusters that are defining the anti-correlation with the ages. In our sample we have taken the abundances of NGC\,3680 \citep{ps18} for calculating the clocks and including them in our sample. This cluster was analyzed using the same line lists and methodology than us and the spectra are also from FEROS. Besides, NGC\,3680 has age $\sim$\,2.0\,GYr and distance $\lesssim$\,1.0\,kpc.

\begin{center}
\begin{table*}
\footnotesize
\caption{Chemical clocks in the clusters IC\,4756, IC\, 4651, NGC\,6940, NGC\,5822, NGC\,6633, NGC\,5316 and NGC\,2682.  The ratios were obtained from the definition [a$/$b]\,=\,$\log\,(N_{\rm a}/N_{\rm b})_{\star}\,-\,\log\,(N_{\rm a}/N_{\rm b})_{\odot}$. Full version is available as supplementary material.}
\label{clockabun7oc}
\begin{tabular}{l|ccccc} \hline \hline
star                 &    [Y$/$Mg]   &      [Y$/$Al]  &        [Y$/$Si]   &	     [Y$/$Ca]  &     [Y$/$Ti]    \\ \hline \hline
IC\,4756-12          &      0.11     &       0.08     &         0.03      &       0.01     &       0.07      \\
IC\,4756-14          &      0.12     &       0.05     &        -0.01	  &       0.08     &       0.20      \\
IC\,4756-28          &      0.01     &      -0.20     &        -0.11      &      -0.02     &       0.08      \\
IC\,4756-38          &      0.18     &       0.10     &         0.07      &       0.08     &       0.12      \\
IC\,4756-42          &      0.15     &       0.06     &         0.00      &       0.03     &       0.10      \\
IC\,4756-44          &      0.23     &       0.10     &         0.03      &       0.01     &       0.09      \\
\end{tabular}
\end{table*}
\end{center}

Thus, it is probable that we are observing only a partial behaviour for the open clusters since the field giants shown the existence of clear trends of [Y$/$Mg], [Y$/$Al], and [Y$/$Si] with age. Then the question that arises is whether the scattering found in the giants is due to the method of obtaining abundances and ages which would hamper the tendencies. This last could be discarded because the field dwarfs, investigated with the same methodology, shown a clear trend with less scattering.

\subsection{Cluster dwarfs and other samples}

The best scenario to study the probable existence of age-abundance relation differences between giants and dwarfs are the open clusters, thus we need to investigate the clusters that were spectroscopically studied from giants and dwarf stars. 

\begin{figure*}
\includegraphics[scale=0.2]{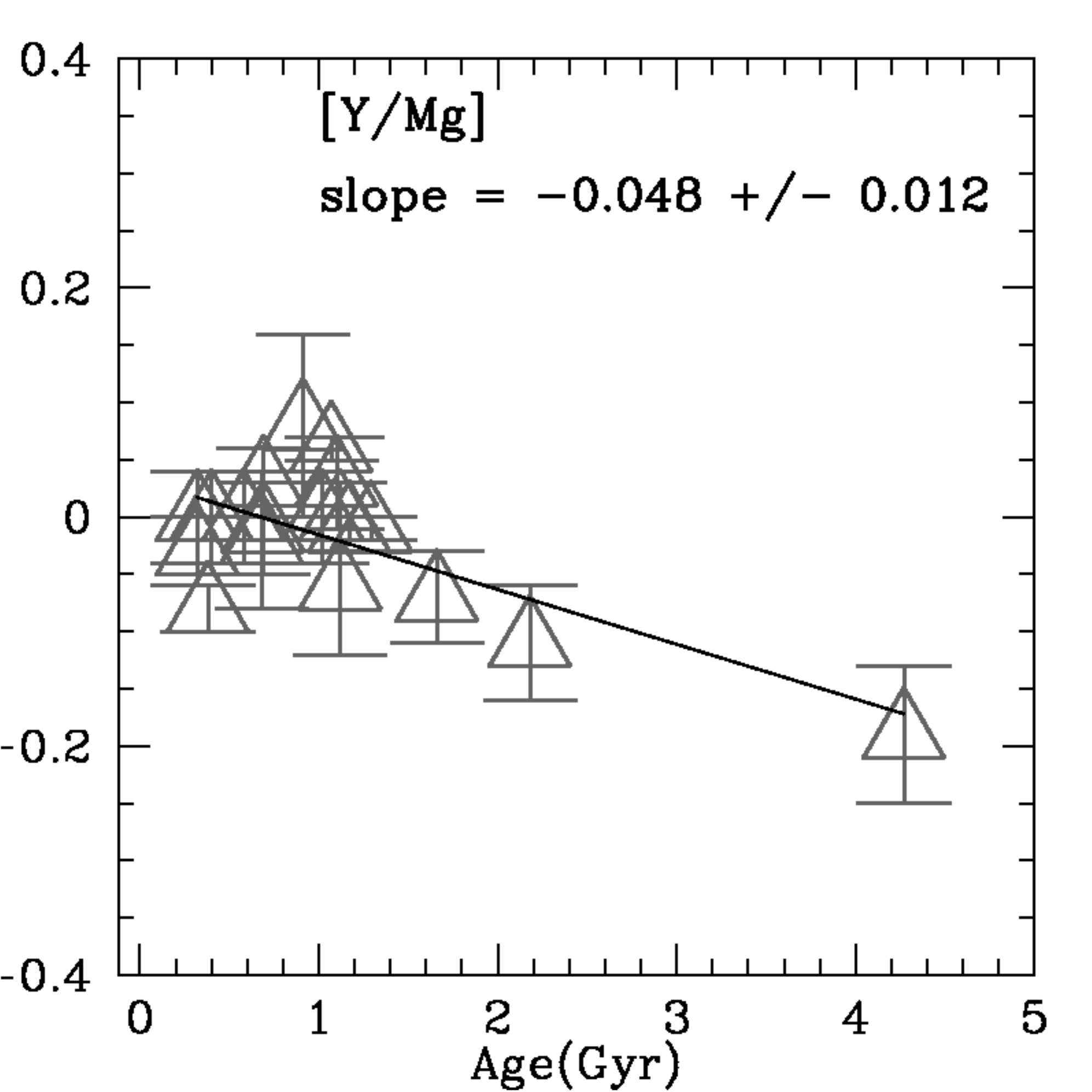}
\includegraphics[scale=0.2]{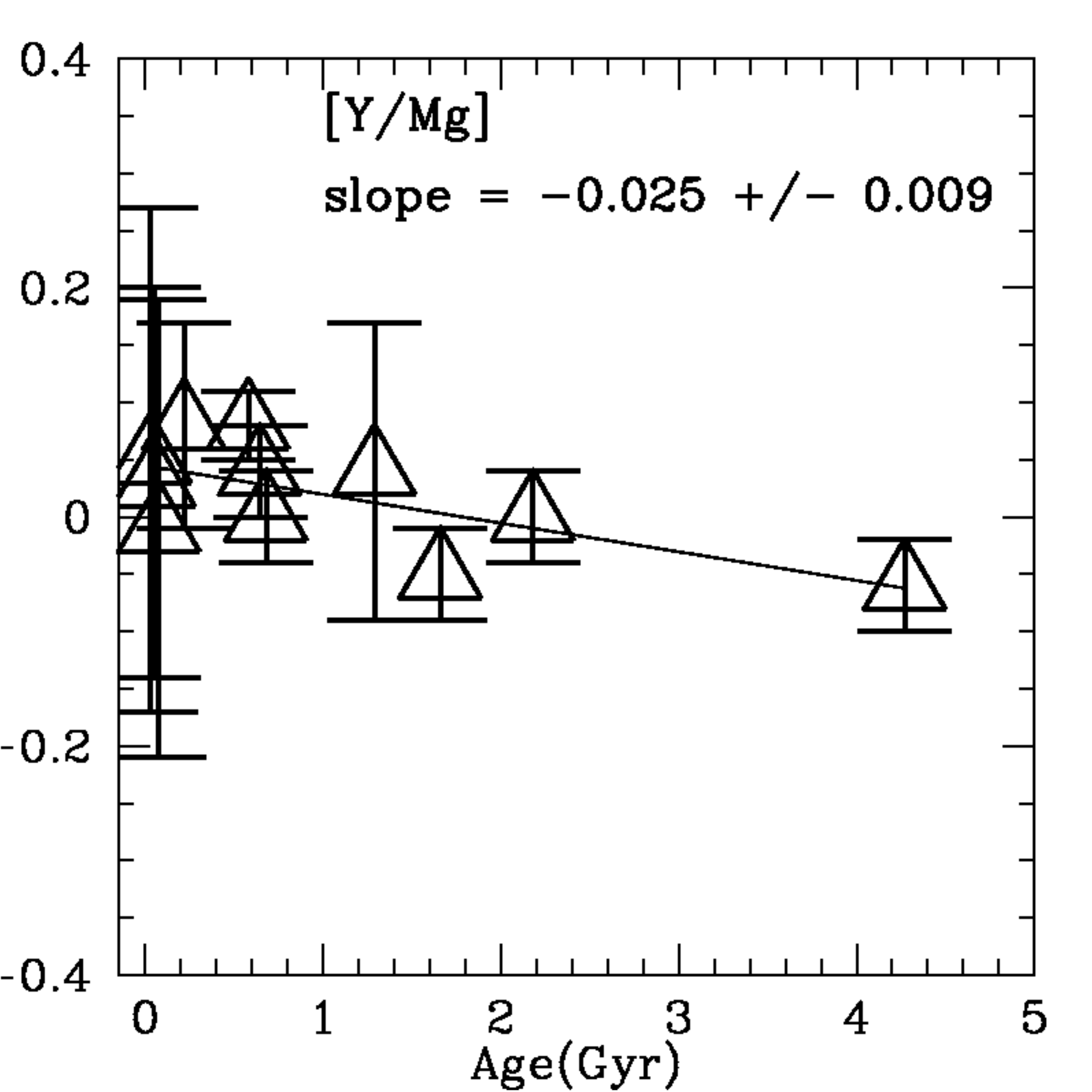}
\includegraphics[scale=0.2]{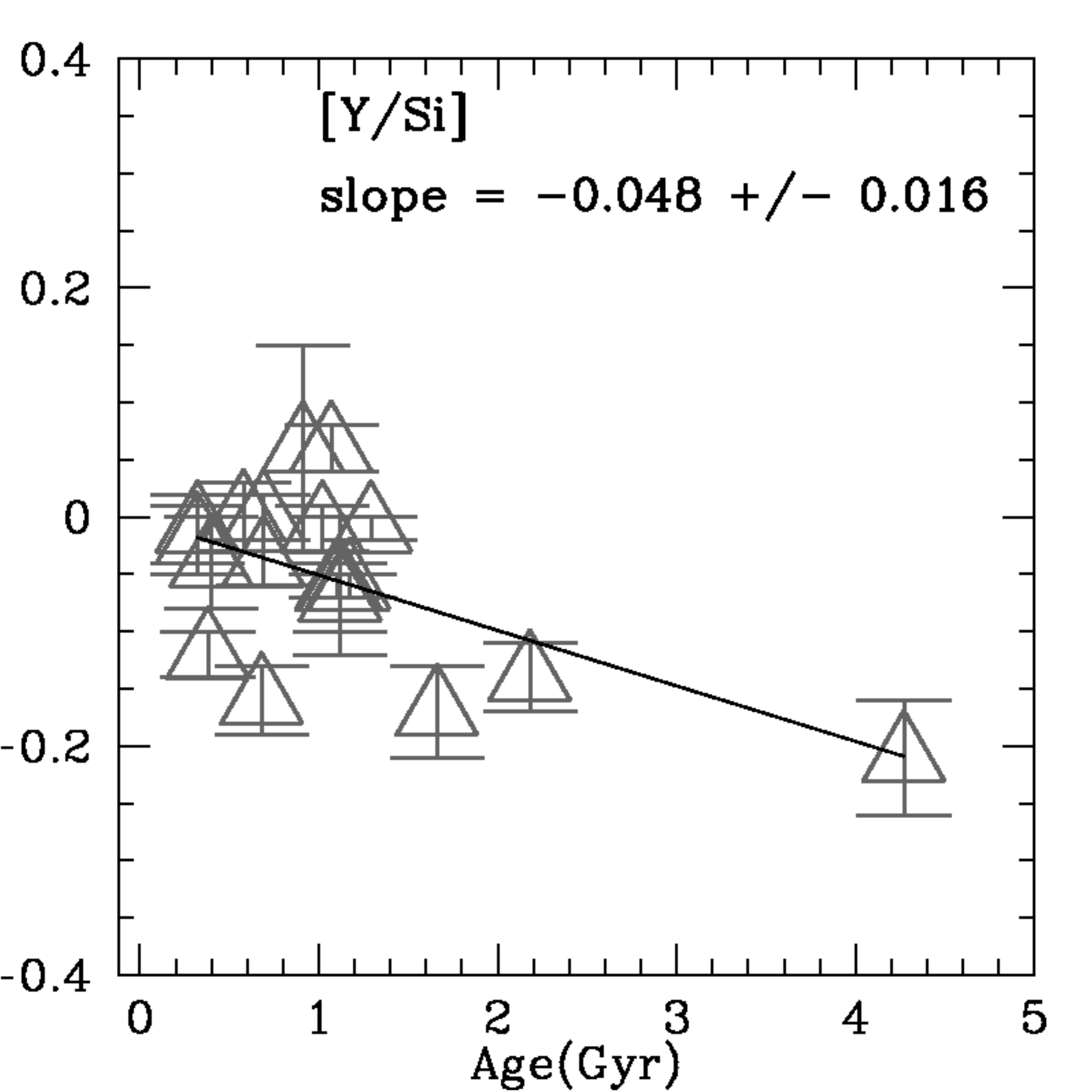}
\includegraphics[scale=0.2]{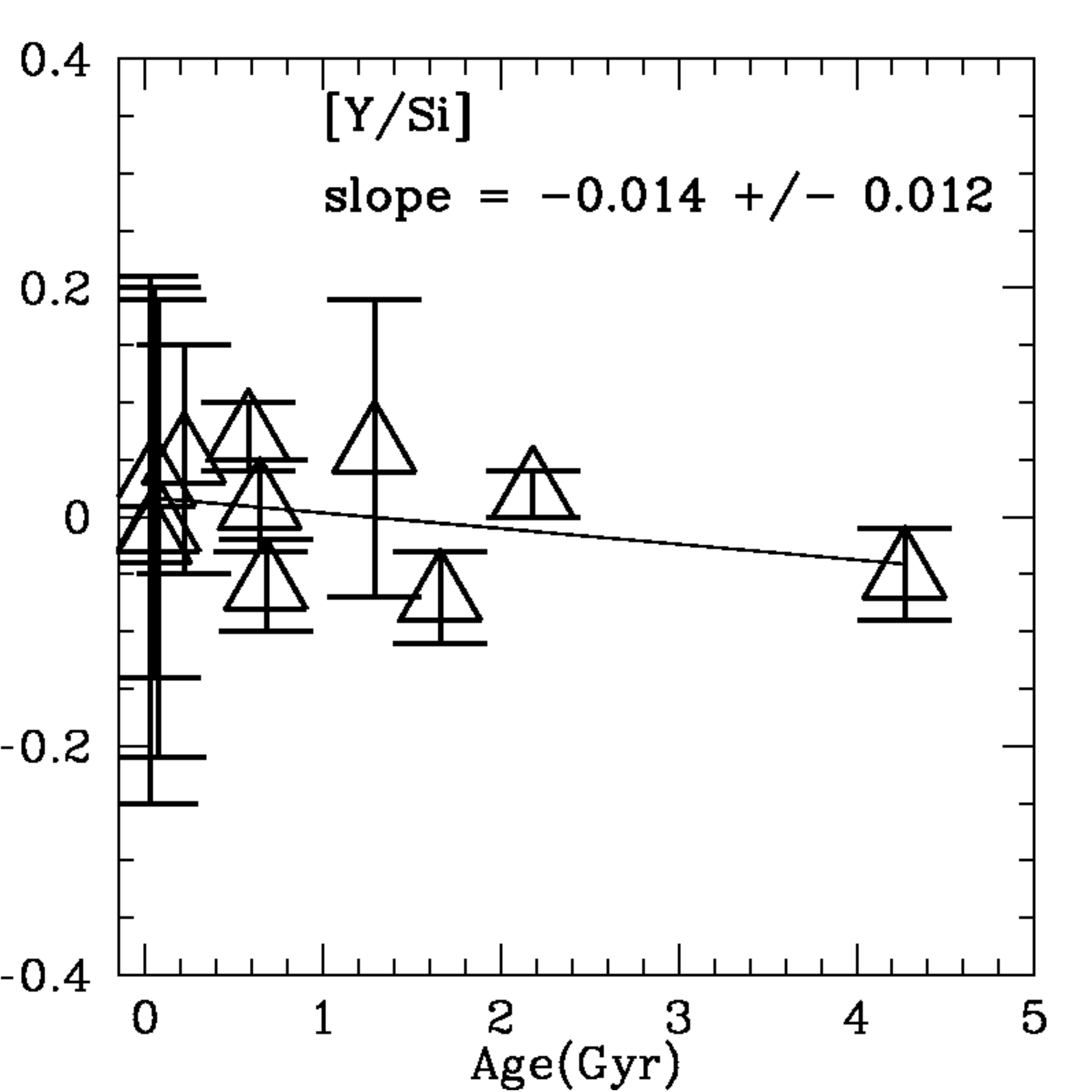}
\includegraphics[scale=0.2]{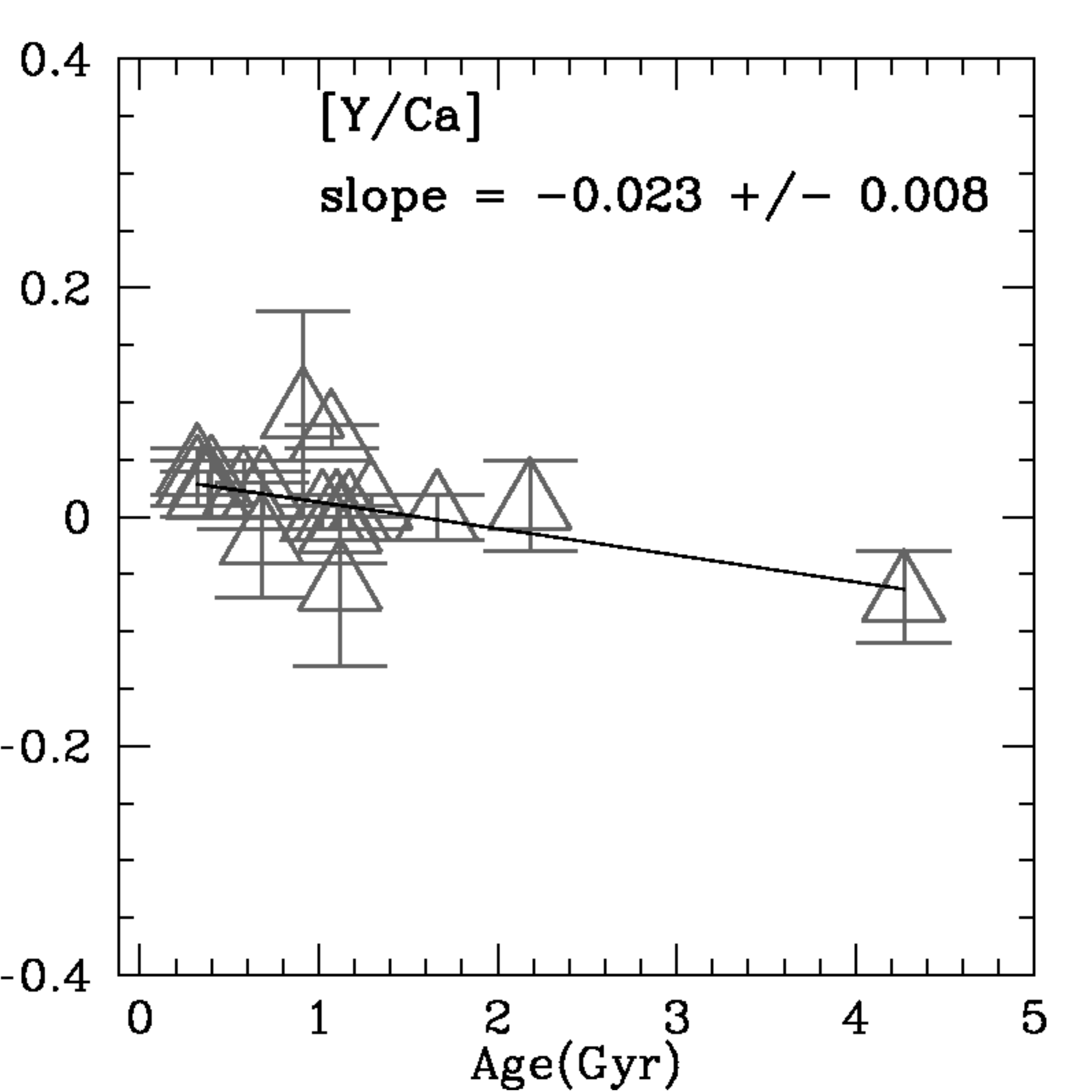}
\includegraphics[scale=0.2]{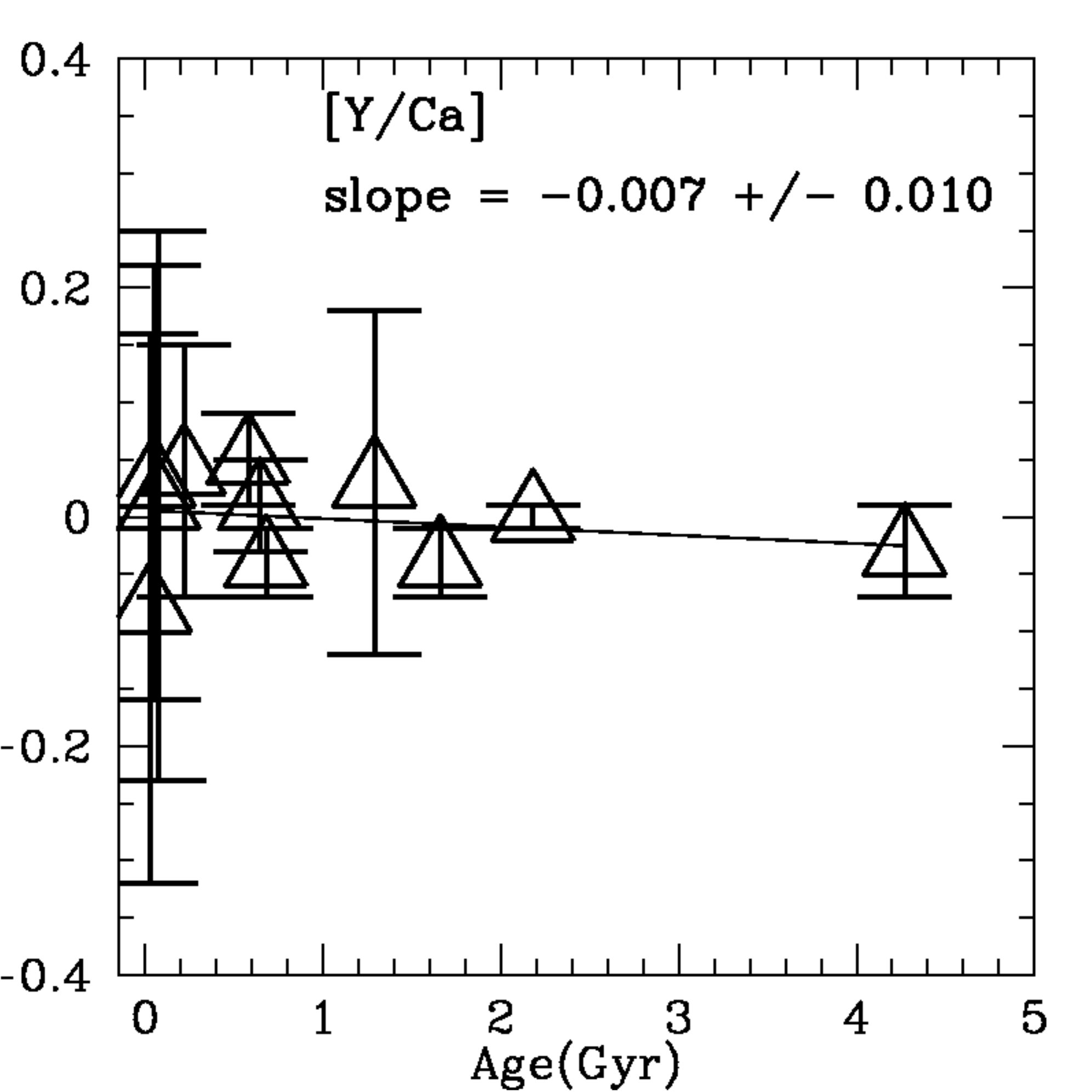}
\includegraphics[scale=0.2]{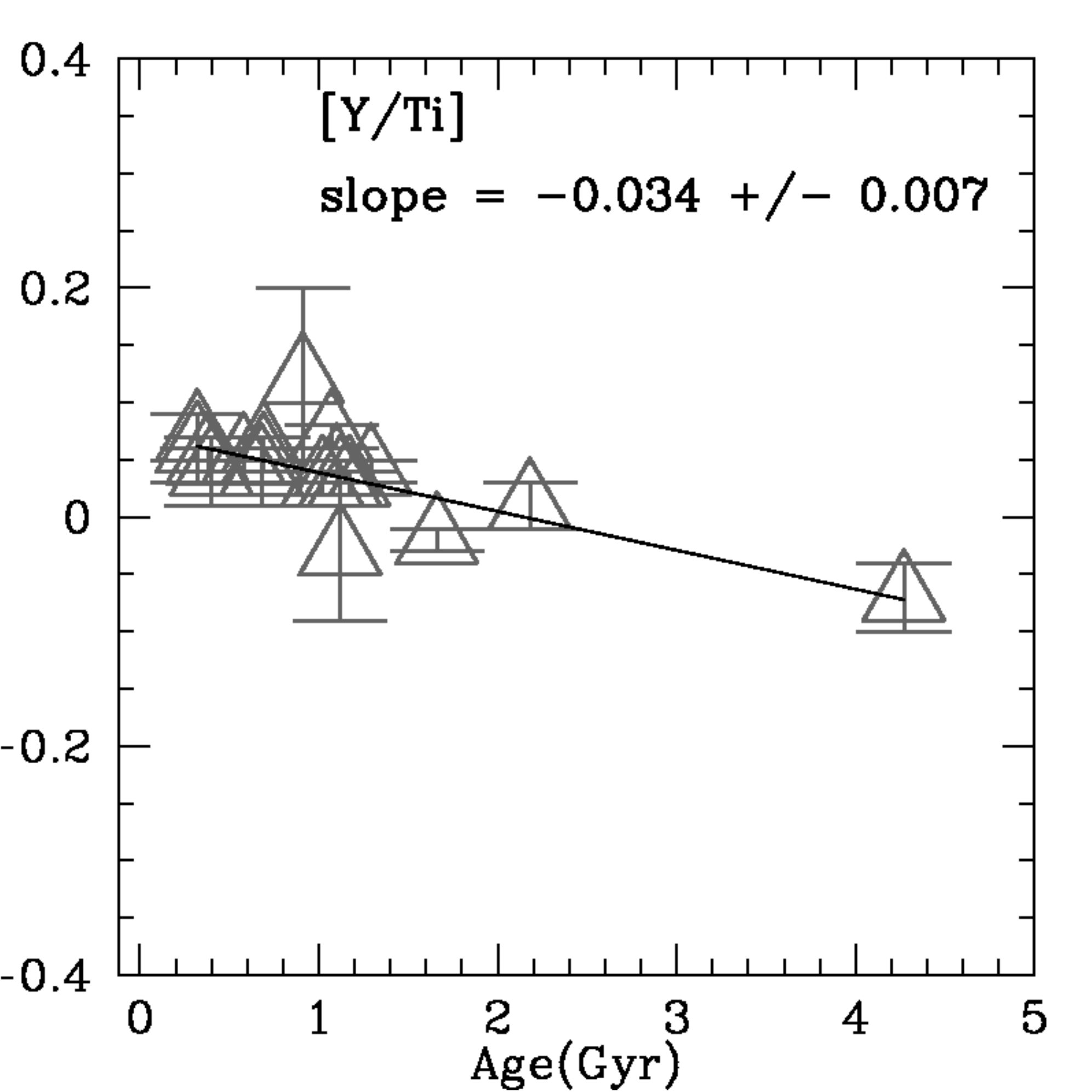}
\includegraphics[scale=0.2]{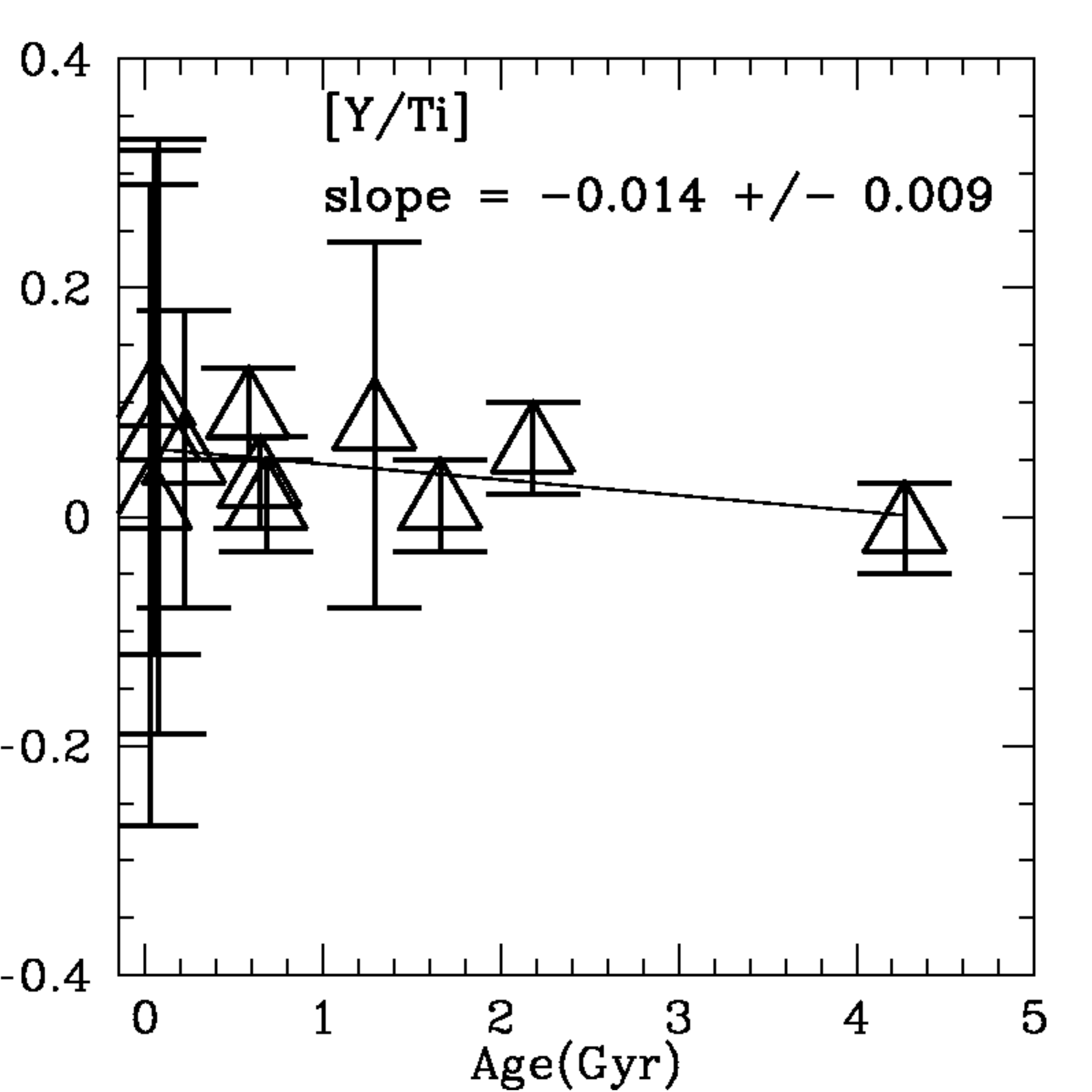}
\caption{Chemical clocks for giant stars (grey open triangles) and dwarfs (black open triangles) in several open clusters studied by \citet{BC2015}.}
\label{ocbc15}
\end{figure*}

On this direction, we have used the table\,8 from \citet{BC2015} to derive these relations which shown in Figure\,\ref{ocbc15}. We observe that the literature tendencies also seem to exist in giants and dwarfs but with different slopes. In opposite direction to the field stars, the four chemical clocks in the sample of \citet{BC2015} show that the cluster giant stars have steeper slopes than the cluster dwarfs.

In the specific case of [Y$/$Mg], our results, \citet{lu15}, and \citet{BC2015} present similar slopes -0.045$\pm$0.023; -0.035$\pm$0.039 and -0.048$\pm$0.012 respectively. Whereas the Reddy's sample presents a almost flat slope. In the cluster dwarfs, this chemical clock present a flatter slope -0.025$\pm$0.009 that is very similar to the field dwarfs.

The case of [Y$/$Si] is complex, our results and \citet{BC2015} have similar slopes -0.057$\pm$0.014 and -0.048$\pm$0.018 respectively. However the other samples present different behaviours: \citet{lu15} has a positive slope 0.015$\pm$0.034 whereas Reddy's sample present a flat. The cluster dwarfs present a flatter slope which agrees with the field dwarfs, ie -0.014$\pm$0.009.

Our relation [Y$/$Ca]\,vs.\,Age is a flat -0.005$\pm$0.019, similar to the flat 0.000$\pm$0.019 of Reddy's sample whereas \citet{lu15} and \citet{BC2015} present similar behaviours with slopes -0.030$\pm$0.029 and -0.023$\pm$0.008 respectively. The cluster dwarfs also present a flat slope in this chemical clock.

Our relation [Y$/$Ti]\,vs.\,Age\,=\,-0.024$\pm$0.019 is similar to the value -0.034$\pm$0.007 from \citet{BC2015} and also similar to -0.043$\pm$0.025 from \citet{lu15}. Whereas Reddy's sample has a flat. This chemical present a flatter slope -0.014$\pm$0.009 in the cluster dwarfs.

In Table\,\ref{gdlit} we see that only seven open clusters were studied using giants and dwarfs, the number of analyzed stars by cluster is also low. However, \citet{luck18} analyzed Melotte\,25 using 42 dwarfs and \citet{BC2015} used 28 giants and 14 dwarfs to analyze NGC\,2682.

There are some differences ($\geqslant$\,0.15\,dex), between giants and dwarfs, that are precisely in these two clusters. Specifically, in the chemical clocks [Y$/$Si] and [Y$/$Al] in Melotte\,25 and in the clock [Y$/$Si] in NGC\,2682. 
Therefore, there is the possibility that by increasing the number of stars by clusters, the differences may increase.

\begin{center}
\begin{table*}
%\centering
\caption{Chemical clocks differences between giants and dwarf stars within the same open clusters. Melotte\,25 was studied by \citet{lu15, luck18} whereas the others were studied by \citet{BC2015}. The N$_{s}$ corresponds to the number of giants and dwarfs studied in each cluster.}
\begin{tabular}{c|ccccc|c} \hline \hline
Cluster        &  [Y$/$Mg]$_{gd}$ &  [Y$/$Al]$_{gd}$   &  [Y$/$Si]$_{gd}$  &  [Y$/$Ca]$_{gd}$  & [Y$/$Ti]$_{gd}$  &  N$_{s}\,(g, d)$ \\  \hline  \hline
Melotte\,25    &   0.05    &	 0.15     &    0.23    &     ---     &   0.07    &     (5, 42)   \\
IC\,4651       &   0.02    &	 ---      &    0.10    &    0.04     &   0.03    &     (6, 2)    \\
IC\,4756       &   0.05    &	 ---      &    0.07    &    0.02     &   0.04    &     (1, 3)    \\
M\,67          &   0.13    &	 ---      &    0.16    &    0.04     &   0.06    &     (28, 14)  \\
NGC\,2447      &   0.14    &	 ---      &    0.13    &    0.11     &   0.12    &     (3, 3)    \\
NGC\,2632      &   0.00    &	 ---      &    0.05    &    0.04     &   0.02    &     (11, 2)   \\
NGC\,3680      &   0.03    &	 ---      &    0.08    &    0.00     &   0.02    &     (2, 3)    \\  \hline
\end{tabular}
\label{gdlit}
\end{table*}
\end{center}

\subsection{Comparisons with literature} \label{abundances_from_lit}

We did comparisons between our ratios [Y$/$Mg], [Y$/$Al], [Y$/$Si], [Y$/$Ca], and [Y$/$Ti] and other authors that done  detailed studies of the same clusters. This means the authors that have the tables, of the analyzed stars, available for star-by-star comparisons. These comparisons are shown in Tables\,\ref{compartab1}, \ref{compartab2} and \ref{compartab3}.

Some stars of the clusters IC\,4651 and NGC\,2682 were also analyzed by \citet{lu15}, we have done a direct comparison with the six stars in common. As it is shown in Table\,\ref{compartab2}, the differences in relation to our values are very low. This is more evidence when we see the mean of the differences. Those similarities are also observed in Figure\,\ref{YallOCRC}, which shows that our results fall along the global behaviour of the field giants of \citet{lu15}.

In Table\,\ref{compartab1}, we see that our results are very similar to the obtained abundances by \citet{ps18} in the cluster NGC\,5822, the quantities $\Delta$[Y$/$Mg], $\Delta$[Y$/$Al], $\Delta$[Y$/$Si], $\Delta$[Y$/$Ca], and  $\Delta$[Y$/$Ti] that represent the mean differences of the abundances have values near to zero. These authors also studied NGC\,3680, this cluster was included in our analysis. NGC\,5822 and NGC\,3680 were studied using the same method than in this work.

The case of NGC\,6940 is similar to NGC\,5822, the mean values of \citet{bocek16} are similar to the found in this work in spite that they did not use the lbl method nor the line lists used by us. \citet{bocek15} also studied NGC\,752 which was normalized to the Juno values.

Also in Table\,\ref{compartab1} we see that IC\,4756 presents notable differences in [Y$/$Mg] and [Y$/$Si] obtained by \citet{bagdo18} and this work. These authors reported lower values and they did not apply the lbl method. Their iron line lists are not the same than us, originating discrepancies in the atmospheric parameters that will also produce differences in the abundance determinations. In the specific case of [Y$/$Mg], these authors only used two absorption lines for obtaining the Mg-abundances in all stars of their sample. In any case, our relations [Y$/$Mg] $\&$ [Y$/$Al]\,$-$age are more near to the literature tendencies than the results of \citet{bagdo18}.

Great differences in NGC\,5316 appear when our results are compared to \citet{draz16}. Our abundance ratios seem to be more congruent in the context of the spectroscopic clocks, the [Y$/$Mg] and [Y$/$Si] from these authors are lightly higher to field giants of \citet{lu15} such as it is observed in the Figure\,\ref{YallOCRC}.

Comparisons with \citet{casali} are shown in Table\,\ref{compartab3}, there are non-negligible differences for the clock [Y$/$Al] in NGC\,6633 $\&$ NGC\,2682 as well as for [Y$/$Si] in NGC\,2682. In the case of NGC\,2682, \citet{casali} reported solar values for the chemical clocks whereas our results also shown values near to the solar but excepting the clocks [Y$/$Al] $\&$ [Y$/$Si] which have presented lower values. A study with more giant stars is necessary to determine if these stars within NGC\,2682 also have solar abundance ratios. In literature there are extensive analysis about M67 but in the majority of these works the number of giant stars is low or simply these stars were not included in the analysis. The open cluster M67 was also analyzed by \citet{slum17} and \citet{onehag}. When comparing with this work and \citet{casali}, we can see that our value [Y$/$Mg]\,=\,-0.07$\pm$0.02\,dex is closer to -0.04$\pm$0.05\,dex found by \citet{onehag}, whereas the results of 0.01$\pm$0.03\,dex \citep{slum17} and 0.00$\pm$0.01\,dex \citep{casali} are very similar. In the case of [Y$/$Al] in NGC\,6633, the interpretation is similar, Our results are within the trend set by the field giants whereas the value reported by \citet{casali} is on the upper limit of the tendency of \citet{lu15}. Unfortunately, \citet{casali} did not specify the studied giant stars in the two clusters, the identifying of the giants would have facilitated the direct comparison.

\begin{center}
\begin{table*}
\footnotesize
\caption{Comparisons star by star $\Delta$[a/b]=[a/b]$_{li}-$[a/b]$_{tw}$, which is the literature value\,$-$\,this work. IC\,4756 compared with  \citet{bagdo18}, NGC\,6940 with \citet{bocek16}, NGC\,5822 with \citet{ps18} and NGC\,5316 with \citet{draz16}. Full version as supplementary material.}
\label{compartab1}
\begin{tabular}{l|ccccc} \hline \hline
Star            &	$\Delta$[Y/Mg] & $\Delta$[Y/Al] & $\Delta$[Y/Si]   & $\Delta$[Y/Ca]  & $\Delta$[Y/Ti]  \\ \hline \hline
IC\,4756-12     &      -0.23       &       -0.22    &     -0.18        &       -0.11     &      -0.03	  \\
IC\,4756-14     &      -0.24       &       -0.13    &     -0.21        &       -0.13     &      -0.01	  \\
IC\,4756-28     &      -0.10       &        0.11    &     -0.21        &       -0.03     &       0.03     \\
IC\,4756-38     &      -0.21       &       -0.02    &     -0.09        &       -0.03     &       0.08     \\
IC\,4756-42     &      -0.28       &       -0.05    &     -0.19        &       -0.09     &       0.01     \\
IC\,4756-44     &      -0.25       &       -0.15    &     -0.10        &       -0.11     &       0.00     \\
\end{tabular}
\end{table*}
\end{center}

\begin{center}
\begin{table*}
\caption{Comparisons between \citet{lu15} and our results.}
\label{compartab2}
\begin{tabular}{c|ccccc} \\ \hline \hline
Star                &	$\Delta$[Y/Mg] & $\Delta$[Y/Al] & $\Delta$[Y/Si] & $\Delta$[Y/Ca] & $\Delta$[Y/Ti]  \\ \hline \hline
NGC\,2682-151       &      -0.05      &      -0.07      &       0.02     &    -0.05       &     0.03       \\
NGC\,2682-244       &       0.01      &       0.03      &       0.05     &    -0.02       &     0.06       \\
IC\,4651-8540       &       0.09      &       0.09      &       0.00     &    -0.05       &    -0.05       \\
IC\,4651-9122       &      -0.06      &      -0.14      &      -0.06     &    -0.09       &    -0.06       \\
IC\,4651-9791       &      -0.11      &      -0.08      &      -0.08     &    -0.06       &    -0.09       \\
IC\,4651-14527      &      -0.05      &      -0.06      &      -0.09     &     0.01       &     0.09       \\ \hline
Mean$\pm\sigma_{n}$ & -0.03$\pm$0.03  &  -0.04$\pm$0.03 & -0.03$\pm$0.02 & -0.04$\pm$0.01 & 0.00$\pm$0.03  \\ \hline
\end{tabular}
\end{table*}
\end{center}

\begin{center}
\begin{table*}
\caption{Comparisons between \citet{casali} and our results.}
\label{compartab3}
\begin{tabular}{c|ccccc} \\ \hline \hline
Cluster    &   $\Delta$[Y/Mg]  &  $\Delta$[Y/Al]  &  $\Delta$[Y/Si]  &  $\Delta$[Y/Ca]  &   $\Delta$[Y/Ti]  \\ \hline \hline
NGC\,6633  &     -0.02         &     0.19         &      0.03        &      0.08        &       0.11        \\
NGC\,2682  &      0.07         &     0.13         &      0.15        &      0.01        &       0.03        \\ \hline
\end{tabular}
\end{table*}
\end{center}

\section{Summary and Conclusions}\label{conclusions}

In this work we have investigated the spectroscopic age indicators [Y$/$Mg], [Y$/$Al], [Y$/$Si, [Y$/$Ca], and [Y$/$Ti] in giant stars. For such purpose, we have done a homogeneous spectroscopic analysis of 50 solar-metallicity giant stars of seven open clusters (NGC\,5316, NGC\,6633, NGC\,5822, IC\,4756, NGC\,6940, IC\,4651, and NGC\,2682). With the goal to generalize our results, we have compared with other homogeneous samples of literature where giants and dwarfs, in the field and open clusters, were investigated. The main results and conclusions that emerge from this work are:

\begin{itemize}

\item From the slopes, correlation coefficients and $p-$values derived in the field dwarf stars, we confirmed that the aforementioned spectroscopic age indicators work in this type of stars. This fact is in agreement with the reported in literature for dwarf stars in the inner disk.

\item For the field giant stars, [Y$/$Mg]- and [Y$/$Al]-age relations present weaker correlations than the ones from dwarf stars, although the slopes are similar. The [Y$/$Si]\,vs.\,age relation shows a correlation coefficient near to the values found in the dwarf stars and a steeper slope. These abundance relations for field giant stars present a remarkable scattering. The ratios [Y$/$Ca] and [Y$/$Ti] have flatter slopes and null correlations with age, indicating that they cannot be considered as age indicators for evolved stars of the inner disk.

%%\item \textbf{To  Atomic diffusion is probably responsible for the differences found in the giant and dwarf stars at the local region. Diffusive effects appear to be present in dwarfs, causing their spectroscopic clocks to decrease.}

\item Our open clusters and the homogeneous literature samples shown that the scattering reported by \citet{casali} to seem to be real independently of the galactic region (7\,$<\,R_{gc}\,<\,$\,8.5\,kpc and/or d\,<\,1\,kpc). This behaviour is observed in the results obtained from giant stars as well as from dwarfs.

\item Despite that the scattering at the early ages, our open clusters shown that the tendency between abundance ratios and ages very probably exist as it has been reported by other authors. This fact is evident for the spectroscopic clocks [Y$/$Mg], [Y$/$Al] and [Y$/$Si] whose slopes -0.045$\pm$0.023, -0.040$\pm$0.015 and -0.057$\pm$0.014, respectively, are close to the literature. Our slope for [Y$/$Ca] is a flat, it agrees with the field giants and other cluster samples.

\item Except for the spectroscopic clock [Y$/$Ca], our cluster stars present steeper slopes than the field stars. This fact could be due to the high number of field stars studied in \citet{lu15, luck18}. Our sample is limited, only two open clusters with ages $>$\,2.0\,GYrs and none cluster with ages $>$\,4.0\,GYr. The clusters NGC\,3680 and NGC\,2682 are the objects that really are defining the anti-correlation for our spectroscopic clocks. However, this situation is very similar to the found in \citet{casali} and \citet{casami21}, in these works, two or three clusters are also defining the anti-correlations. These facts reinforce the need to analyze the older open clusters. 

%\item \textbf{For the spectroscopic age indicators analyzed using open clusters, there are high scattering at 7 $< R_{gc} < $ 8.5\,kpc and R$_{gc} > $ 8.5. In this sense, our open clusters of the solar neighbourhood shown similar behaviour to the field giants, those localized in the outer disk seem to present probable trends between their abundance ratios and ages but with high scattering. We have only to M67 as the old open cluster that is defining the trend, therefore the derived trends for outer disk are not conclusive.}

\item Because of the behaviour found in Melotte\,25 and NGC\,2682, it is also need to study more giant stars and dwarfs by cluster to confirm/refuse the differences between these two type of stars.

\item From this work and the literature, we seen that several relations between abundance ratios and ages seem to exist in the Galactic disk when giants and dwarfs are analyzed. This fact point to the probable non-universality of the spectroscopic age indicators.

\item The abundance ratios [Sr$/$Fe], [Ba$/$Fe], [s$/$Fe] and [Eu$/$Fe] derived in our open clusters follow the Galactic disk trend reported in other clusters and field stars of literature.

\end{itemize}

\section*{Acknowledgements}

The authors would like to acknowledge to Coordena\c c\~ao de Aperfei\c coamento de Pessoal de Nível Superior - Brasil (CAPES) - to be the main financier of this work. As well as to Conselho Nacional de Desenvolvimento Científico e Tecnológico CNPq for its partial financial support. The authors also want to acknowledge to the anonymous referee for his/her important discussions that help us to improve this work. Data was obtained from the ESO Science Archive Facility under request numbers 216546 and 230944 and username ojksantrich. This work presents results from the European Space Agency (ESA) space mission \textit{Gaia}. \textit{Gaia} data are being processed by the \textit{Gaia} Data Processing and Analysis Consortium (DPAC). Funding for the DPAC is provided by national institutions, in particular the institutions participating in the \textit{Gaia} MultiLateral Agreement (MLA). The \textit{Gaia} mission website is https://www.cosmos.esa.int/gaia. The \textit{Gaia} archive website is https://archives.esac.esa.int/gaia.

\section*{Data Availability}

The authors confirm that the all data supporting the findings of this study are openly available in ESO/archive at http://archive.eso.org/cms.html and can be accessed in free way. As well as there are available data within the article and/or its supplementary material.

%\newpage

%%%%%%%%%%%%%%%%%%%%%%%%%%%%%%%%%%%%%%%%%%%%%%%%%%
\clearpage
\onecolumn

%\appendix
%\section{TABLES}
%\input{sampleA1.tex}
%\input{IC4756.tex}
%\input{M67.tex} 
%\input{NGC5316.tex} 
%\input{NGC5822.tex}
%\input{NGC6633.tex}
%\input{NGC6940.tex}
%\twocolumn

%%%%%%%%%%%%%%%%%%%%%%%%%%%%%%%%%%%%%%%%%%%%%%%%%%
% Don't change these lines
\bsp	% typesetting comment
\label{lastpage}
\end{document}